\documentclass[twocolumn,tighten]{aastex63}
\hypersetup{linkcolor=blue,citecolor=blue,filecolor=blue,urlcolor=blue}
\usepackage{amsmath}
\usepackage{natbib}
\usepackage{enumitem}

\newcommand{\jtwo}{\object[2MASS J22132050-5137385]{J2213$-$5137}}
\newcommand{\jtwolong}{\object[2MASS J22132050-5137385]{2MASS~J22132050$-$5137385}}
\newcommand{\cainstar}{\object[2MASS J15213995-3538094]{J1521$-$3538}}
\newcommand{\cainstarlong}{\object[2MASS J15213995-3538094]{2MASS~J15213995$-$3538094}}

\newcommand{\hdone}{\object[HD 184266]{HD~184266}}
\newcommand{\hdtwo}{\object[HD 222925]{HD~222925}}

\newcommand{\loggf}{\mbox{$\log(gf)$}}
\newcommand{\kmsec}{\mbox{km~s$^{\rm -1}$}}
\newcommand{\logg}{\mbox{log~{\it g}}}
\newcommand{\msun}{\mbox{$M_{\odot}$}}
\newcommand{\teff}{\mbox{$T_{\rm eff}$}}
\newcommand{\vt}{\mbox{$v_{\rm t}$}}
\newcommand{\rpro}{\mbox{{\it r}-process}}
\newcommand{\spro}{\mbox{{\it s}-process}}

\newcommand{\rettwolong}{\object[NAME RETICULUM II]{Reticulum~II}}

\newcommand{\fornax}{\object[NAME FORNAX DSPH]{Fornax}}

\newcommand{\rtwo}{\mbox{\it r}-II}
\newcommand{\rthree}{\mbox{\it r}-III}
\newcommand{\logeps}[1]{$\log\varepsilon$(#1)}
\newcommand{\mualpha}{\mu_{\alpha*}}
\newcommand{\mudelta}{\mu_\delta}
\newcommand{\agama}{\texttt{AGAMA}}
\newcommand{\deltaeufe}{$\delta_{\rm EuFe}$}
\def\vector#1{\mbox{\boldmath $#1$}}

\shorttitle{The r-process-enhanced star J2213--5137}
\shortauthors{Roederer et al.}
\accepted{for publication in the Astrophysical Journal}

\begin{document}

\title{%
The $R$-Process Alliance:\
2MASS~J22132050$-$5137385,
the Star with the Highest-known \textit{r}-process Enhancement
at [Eu/Fe] = +2.45%
\footnote{%
This paper includes data gathered with the 6.5~meter 
Magellan Telescopes located at Las Campanas Observatory, Chile.}}

\author[0000-0001-5107-8930]{Ian U.\ Roederer}
\affiliation{%
Department of Physics, North Carolina State University,
Raleigh, NC 27695, USA}
\affiliation{%
Joint Institute for Nuclear Astrophysics -- Center for the
Evolution of the Elements (JINA-CEE), USA}
\email{Email:\ iuroederer@ncsu.edu}

\author[0000-0003-4573-6233]{Timothy C.\ Beers}
\affiliation{%
Department of Physics and Astronomy, University of Notre Dame, 
Notre Dame, IN 46556, USA}
\affiliation{%
Joint Institute for Nuclear Astrophysics -- Center for the
Evolution of the Elements (JINA-CEE), USA}

\author[0000-0001-6924-8862]{Kohei~Hattori}
\affiliation{National Astronomical Observatory of Japan, 2-21-1 Osawa, 
Mitaka, Tokyo 181-8588, Japan}
\affiliation{The Institute of Statistical Mathematics, 10-3 Midoricho, 
Tachikawa, Tokyo 190-8562, Japan}

\author[0000-0003-4479-1265]{Vinicius M.\ Placco}
\affiliation{%
NSF's NOIRLab, 
Tucson, AZ 85719, USA}

\author[0000-0001-6154-8983]{Terese T.\ Hansen}
\affiliation{%
Department of Astronomy, 
Stockholm University, 
SE-106 91 Stockholm, Sweden}

\author[0000-0002-8504-8470]{Rana Ezzeddine}
\affiliation{%
Department of Astronomy, University of Florida, Bryant Space Science Center,  
Gainesville, FL 32611, USA}
\affiliation{%
Joint Institute for Nuclear Astrophysics -- Center for the
Evolution of the Elements (JINA-CEE), USA}

\author[0000-0002-2139-7145]{Anna Frebel}
\affiliation{%
Department of Physics and Kavli Institute for Astrophysics and Space Research, 
Massachusetts Institute of Technology, 
Cambridge, MA 02139, USA}
\affiliation{%
Joint Institute for Nuclear Astrophysics -- Center for the
Evolution of the Elements (JINA-CEE), USA}

\author[0000-0002-5463-6800]{Erika M.\ Holmbeck}
\affiliation{%
Carnegie Observatories,
Pasadena, CA 91101, USA}
\affiliation{%
Joint Institute for Nuclear Astrophysics -- Center for the
Evolution of the Elements (JINA-CEE), USA}

\author[0000-0002-5095-4000]{Charli M.\ Sakari}
\affiliation{%
Department of Physics and Astronomy, San Francisco State University,
San Francisco, CA 94132, USA}

\begin{abstract}

We present stellar parameters and chemical abundances of
47~elements detected in the
bright ($V = 11.63$) 
very metal-poor ([Fe/H] = $-2.20 \pm 0.12$) star
\mbox{2MASS~J22132050$-$5137385}.
We observed this star 
using the Magellan Inamori Kyocera Echelle spectrograph
as part of ongoing work by the \textit{R}-Process Alliance.
The spectrum of \mbox{2MASS~J22132050$-$5137385}
exhibits unusually strong lines of elements
heavier than the iron group, and
our analysis reveals that 
these elements were produced by 
rapid neutron-capture (\rpro) nucleosynthesis.
We derive a europium enhancement,
[Eu/Fe] = $+2.45 \pm 0.08$, 
that is higher than any other 
\rpro-enhanced star known at present.
This star is only the eighth
\rpro-enhanced star where
both thorium and uranium are detected, and
we calculate the age of the \rpro\ material,
$13.6 \pm 2.6$~Gyr, from the radioactive decay of these isotopes.
This star contains relatively large enhancements of
elements that may be produced as transuranic fission fragments, and
we propose a new method using this characteristic
to assess the \rpro\ yields and gas dilution
in samples of \rpro-enhanced stars.
We conclude that 
\mbox{2MASS~J22132050$-$5137385}
exhibits a high level of
\rpro\ enhancement because it formed 
in an environment
where the \rpro\ material was less diluted than average.
Assuming a canonical baryonic minihalo mass of $10^{6}$~\msun\
and a 1\% metal retention rate,
this star formed in a cloud of 
only $\sim 600$~\msun.


\end{abstract}

\keywords{%
Nucleosynthesis (1131);
R-process (1324);
Stellar abundances (1577)
}

\section{Introduction}
\label{sec:intro}

\setcounter{footnote}{12}

Understanding the origins of elements across the periodic table
remains a major challenge of modern astrophysics.
Heavy elements produced by the 
rapid neutron-capture process, or \rpro\ \citep{cowan21},
are not distributed evenly
throughout the Milky Way and Local Group of galaxies.
The enhancement of \rpro\ material relative to the stellar metallicity
is often quantified through the 
[Eu/Fe] ratio\footnote{%
The ratio of the abundances of elements 
europium (Eu, $Z = 63$) and 
iron (Fe, $Z = 26$) relative to the
Solar ratio, [Eu/Fe], is defined as
$\log_{10} (N_{\rm Eu}/N_{\rm Fe}) - \log_{10} (N_{\rm Eu}/N_{\rm Fe})_{\odot}$.
We adopt analogous notation for other abundance ratios.
}.
Among the Sun and other metal-rich stars in the Milky Way thin disk,
there is a small range in [Eu/Fe] of
about 0.3--0.4~dex (a factor of 2.0--2.5;
e.g., \citealt{battistini16,delgadomena17,guiglion18}).
In contrast, metal-poor stars
exhibit a large range in [Eu/Fe] that
varies by 3--4 orders of magnitude
(a factor of 1,000--10,000) 
across the Milky Way halo and its dwarf galaxies
(e.g., \citealt{mcwilliam98,ji20,li22}).

Stars with the highest levels of \rpro\ enhancement,
defined empirically as exhibiting 
[Eu/Fe] ratios $> +0.7$ \citep{holmbeck20},
are frequently referred to as \rtwo\ stars.
More than 200 such stars have been 
confirmed in the Milky Way
over the last 30~years,
to the best of our knowledge of the literature.
The star with the highest known level of \rpro\ enhancement,
\cainstarlong\ (hereafter \cainstar),
with [Eu/Fe] = $+2.23 \pm 0.12$,
was studied in detail by \citet{cain20}.
That study proposed a new class, \rthree,
to designate \rpro-enhanced stars with [Eu/Fe] $> +2$.

\cainstarlong\
shares the extreme tail of the distribution ([Eu/Fe] $> +1.7$)
with 
five other \rpro-enhanced stars known in the Milky Way
\citep{frebel07he,aoki10,jacobson15smss,ezzeddine20,hansen21}
and several stars in the \rettwolong\ dwarf galaxy
\citep{ji16ret2,ji23ret2,roederer16b,hayes23}.
Similarly, 
the \object[name fornax dwarf galaxy]{Fornax}
dwarf galaxy contains several
metal-poor stars that are extremely enriched in
\rpro\ material, [Eu/H] $> +0.0$ 
(enhanced to [Eu/Fe] $\gtrsim +1.3$ with [Fe/H] $\gtrsim -1.3$;
\citealt{reichert21fnx}).
All of these stars are members of 
dwarf galaxies,
kinematic substructures associated with disrupted dwarf galaxies
\citep{roederer18d,yuan20,gudin21,hattori23,shank23},
or stellar streams \citep{hansen21}.
Many theoretical investigations exploring
\rpro\ enrichment in dwarf galaxies
indicate that low-rate and high-yield \rpro\ sites
can generate extreme \rpro\ enhancements in stars
(e.g., \citealt{beniamini16b,ojima18,molero21,hirai22,cavallo23}).

The \textit{R}-Process Alliance (RPA)
has been conducting high-resolution spectroscopic observations
of metal-poor stars to identify and characterize 
several hundred new \rpro-enhanced stars
\citep{hansen18,sakari18north,ezzeddine20,holmbeck20}.
This manuscript details our discovery, analysis, and interpretation
of the nature of one extreme \rpro-enhanced star.
We observed the star \jtwolong\ (hereafter \jtwo)
as part of the RPA activities (Section~\ref{sec:data}).
We recognized the peculiar nature of this star's optical spectrum,
which exhibits hundreds of unusually strong lines of 
singly ionized lanthanide elements (Section~\ref{sec:inspect}).
Our analysis (Section~\ref{sec:analysis})
reveals that this star exhibits the highest level
of \rpro\ enhancement known to date (Section~\ref{sec:results}).
We also detect the radioactive actinide elements
thorium and uranium in its spectrum
(Section~\ref{sec:uranium}),
marking only the eighth time that both elements
have been detected together in an \rpro-enhanced star.
We discuss the 
kinematics of this extreme star 
and the circumstances that may have led to
its formation
(Sections~\ref{sec:dilution} and \ref{sec:dwarfgalaxy}).
Finally, we summarize our conclusions about the 
uncommon nature of this star
(Section~\ref{sec:conclusions}).

\section{Data}
\label{sec:data}

\subsection{Observations}
\label{sec:obs}

Table~\ref{tab:data} lists the basic data for \jtwo.
\jtwo\ was identified as a metal-poor star
in the fourth data release (DR4) from the
RAdial Velocity Experiment (RAVE; \citealt{kordopatis13rave}).
The RAVE DR4 stellar parameter pipeline returned
effective temperature (\teff) of $5870 \pm 158$~K,
log of the surface gravity (\logg) of $3.27 \pm 0.45$ in cgs units, and
metallicity ([Fe/H]) of $-1.63 \pm 0.12$
for \jtwo.
This star was not included in the initial set of
medium-resolution spectroscopic followup of RAVE
targets conducted by the RPA \citep{placco18}.
Nevertheless, that analysis revealed that the
published RAVE stellar parameter estimates
were sufficiently reliable to 
bypass collecting medium-resolution spectroscopic followup for most stars
and proceed directly to high-resolution spectroscopic followup.

\begin{deluxetable*}{lcccc}
\tablecaption{Properties of J2213$-$5137
\label{tab:data}}
\tablewidth{0pt}
\tabletypesize{\scriptsize}
\tablehead{
\colhead{Quantity} &
\colhead{Symbol} &
\colhead{Value} &
\colhead{Units} &
\colhead{Reference}
}
\startdata
\hline\hline
\multicolumn{5}{c}{Astrometric Properties} \\
\hline
Right ascension           & $\alpha$ (J2000)    & 22:13:20.49            & hh:mm:ss.ss   & Gaia DR3 \citep{gaia21dr3astrometry} \\
Declination               & $\delta$ (J2000)    & $-$51:37:38.7          & dd:mm:ss.s    & Gaia DR3 \citep{gaia21dr3astrometry} \\
Galactic longitude        & $\ell$              & 342.4                  & degrees       & Gaia DR3 \citep{gaia21dr3astrometry} \\
Galactic latitude         & $b$                 & $-$51.6                & degrees       & Gaia DR3 \citep{gaia21dr3astrometry} \\
Proper motion ($\alpha$)  & $\mualpha$          & 7.012 $\pm$ 0.017      & mas yr$^{-1}$ & Gaia DR3 \citep{gaia21dr3astrometry}\\
Proper motion ($\delta$)  & $\mudelta$          & $-$17.524 $\pm$ 0.019  & mas yr$^{-1}$ & Gaia DR3 \citep{gaia21dr3astrometry}\\
Parallax                  & $\varpi$            & 0.4647 $\pm$ 0.0186    & mas           & Gaia DR3 \citep{gaia21dr3astrometry} \\
Distance                  & $D$                 & 2.01$^{+0.08}_{-0.10}$ & kpc           & \citet{bailerjones21} \\
Galactocentric coordinates (Cartesian) & ($x$, $y$, $z$) & ($-6.99^{+0.05}_{-0.04}$, $-0.38^{+0.01}_{-0.01}$, $-1.58^{+0.06}_{-0.06}$)
                                                                         & kpc           & this study \\
Galactocentric distance   & $R$                 & $7.00^{+0.04}_{-0.05}$ & kpc           & this study \\
\hline\hline
\multicolumn{5}{c}{Kinematic Properties} \\
\hline
Radial velocity           & $v_{r}$             & 120.59 $\pm$ 1.36      & \kmsec        & RAVE DR5 \citep{kunder17}\tablenotemark{a} \\
Radial velocity           & $v_{r}$             & 119.12 $\pm$ 0.47      & \kmsec        & Gaia DR3 \citep{gaia22dr3rv} \\
Radial velocity           & $v_{r}$             & 117.5 $\pm$ 0.3        & \kmsec        & this study\tablenotemark{b} \\
Galactocentric velocity (Cartesian) & ($v_{x}$, $v_{y}$, $v_{z}$) & ($32.1^{+1.8}_{-1.9}$, $52.1^{+6.2}_{-6.7}$, $-80.1^{+0.3}_{-0.3}$) 
                                                                         & \kmsec        & this study \\
Galactocentric velocity (cylindrical) & ($v_{R}$, $v_{z}$, $v_{\phi}$) & ($-34.9^{+2.2}_{-2.0}$, $-80.1^{+0.3}_{-0.3}$, $50.3^{+6.2}_{-6.7}$)
                                                                         & \kmsec        & this study \\
Integrals of motion       & ($J_{r}$, $J_{z}$, $J_{\phi}$) & ($332.9^{+18.5}_{-16.8}$, $174.2^{+4.2}_{-4.1}$, $352.0^{+45.8}_{-48.8}$)
                                                                         & kpc~\kmsec    & this study \\
Orbital pericenter        & $r_{\rm peri}$      & $1.12^{+0.15}_{-0.15}$ & kpc           & this study \\
Orbital apocenter         & $r_{\rm apo}$       & $7.20^{+0.04}_{-0.04}$ & kpc           & this study \\
Maximum orbital distance from Galactic midplane &$|z_{\rm max}|$& $4.04^{+0.14}_{-0.13}$ & kpc & this study \\
Orbital eccentricity      & $e$                 & $0.73^{+0.03}_{-0.03}$ & \nodata       & this study \\
\hline\hline
\multicolumn{5}{c}{Photometric Properties} \\
\hline
$NUV$ magnitude           & $NUV$               & 15.748 $\pm$ 0.012     & mag           & GALEX \citep{bianchi17} \\
$B$ magnitude             & $B$                 & 12.223 $\pm$ 0.008     & mag           & APASS \citep{munari14} \\
$V$ magnitude             & $V$                 & 11.630 $\pm$ 0.02      & mag           & APASS \citep{munari14} \\
$G$ magnitude             & $G$                 & 11.446 $\pm$ 0.003     & mag           & Gaia DR3 \citep{gaia21dr3phot} \\
$BP$ magnitude            & $G_{BP}$            & 11.788 $\pm$ 0.003     & mag           & Gaia DR3 \citep{gaia21dr3phot} \\
$RP$ magnitude            & $G_{RP}$            & 10.928 $\pm$ 0.004     & mag           & Gaia DR3 \citep{gaia21dr3phot} \\
$J$ magnitude             & $J$                 & 10.311 $\pm$ 0.026     & mag           & 2MASS \citep{cutri03} \\
$H$ magnitude             & $H$                 &  9.949 $\pm$ 0.025     & mag           & 2MASS \citep{cutri03} \\
$K_{s}$ magnitude         & $K_{s}$             &  9.863 $\pm$ 0.023     & mag           & 2MASS \citep{cutri03} \\
Color excess              & $E(B-V)$            & 0.017 $\pm$ 0.01       & mag           & \citet{schlafly11} \\
Bolometric correction     & $BC_{V}$            & $-$0.31 $\pm$ 0.04     & mag           & \citet{casagrande14c} \\
\hline\hline
\multicolumn{5}{c}{Model Atmosphere} \\
\hline
Effective temperature     & \teff               & 5509 $\pm$ 43 (stat.) $\pm$ 72 (sys.) & K & this study \\
Log of surface gravity    & \logg               & 2.28 $\pm$ 0.06        & (cgs)         & this study \\
Microturbulent velocity   & \vt                 & 2.25 $\pm$ 0.10        & \kmsec        & this study \\
Metallicitiy              & [Fe/H]              & $-$2.20 $\pm$ 0.12     & \nodata       & this study \\
\enddata      
\tablenotetext{a}{%
2009/08/20}
\tablenotetext{b}{%
Calculated from three separate MIKE observations:\
RV = $117.5 \pm 0.7$~\kmsec\
on 2021/12/05 at UT = 00:33;
RV = $117.8 \pm 0.7$~\kmsec\
on 2022/07/22 at UT = 05:45; and
RV = $117.3 \pm 0.7$~\kmsec\ 
on 2022/08/18 at UT = 04:25.
}
\end{deluxetable*}

\begin{figure*}
\begin{center}
\includegraphics[angle=0,width=4.9in]{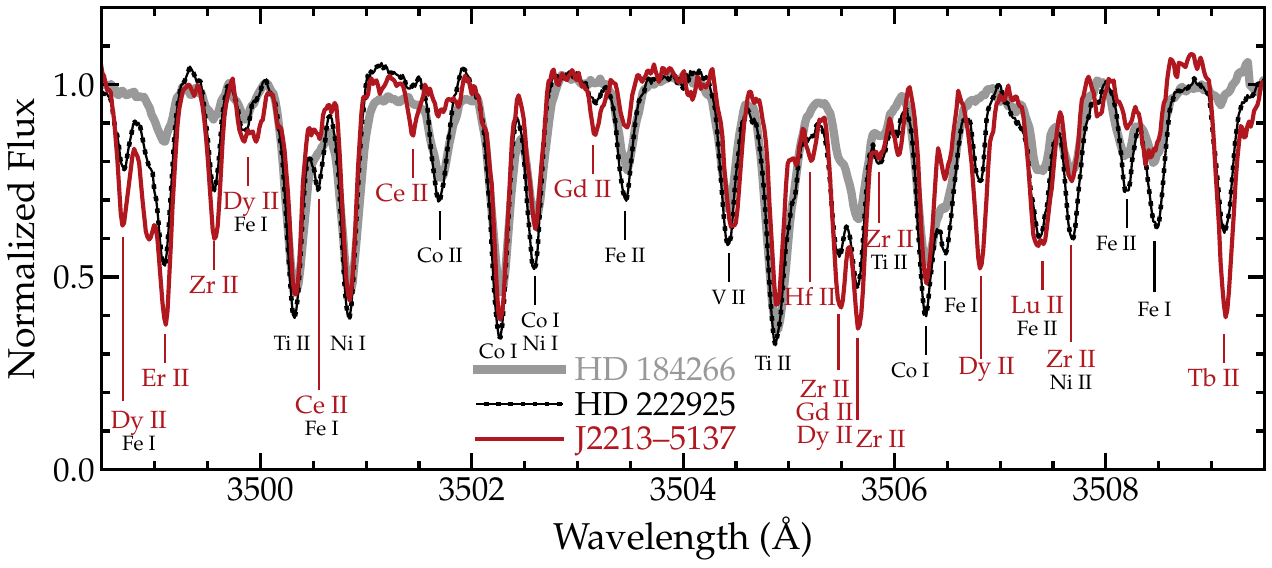}
\vspace*{0.1in}
\includegraphics[angle=0,width=4.9in]{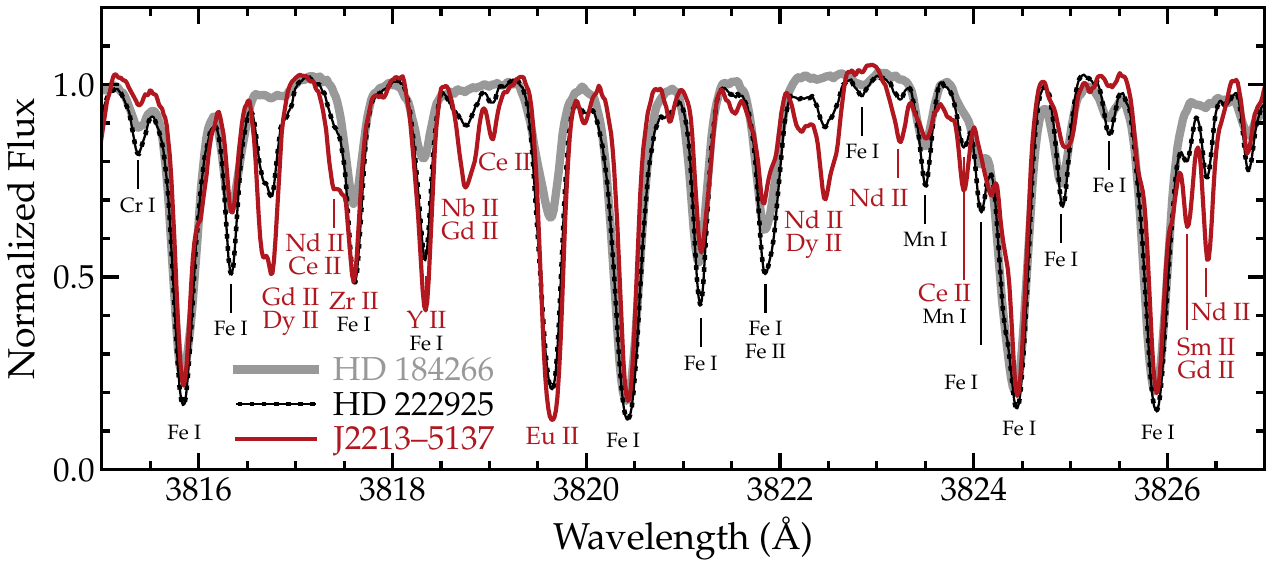}
\vspace*{0.1in}
\includegraphics[angle=0,width=4.9in]{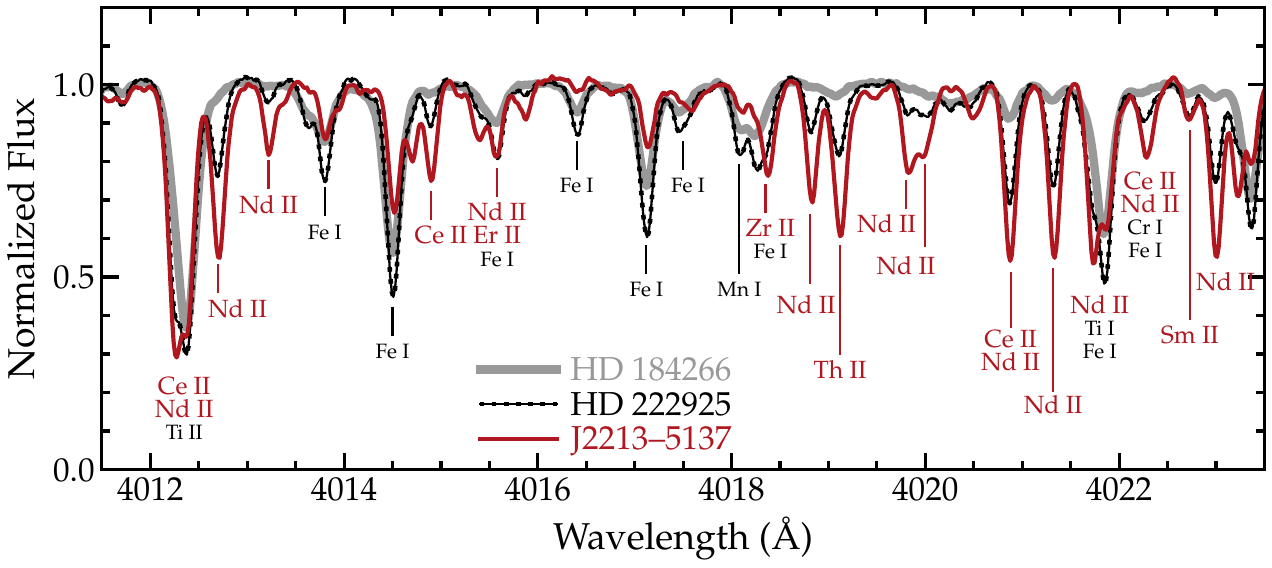}
\end{center}
\caption{
\label{fig:specplot}
Comparison of the spectra of three stars with similar \teff\ 
but contrasting levels of \rpro\ enhancement.
\hdone\ has 
\teff\ = $5580 \pm 102$~K, 
[Fe/H] = $-1.80 \pm 0.08$, and
[Eu/Fe] = $+0.29 \pm 0.16$
\citep{roederer14c,roederer18c}.
\hdtwo\ has
\teff\ = $5636 \pm 103$~K,
[Fe/H] = $-1.46 \pm 0.10$, and
[Eu/Fe] = $+1.32 \pm 0.08$
\citep{roederer18c,roederer22a}.
\jtwo\ has
\teff = $5509 \pm 72$~K,
[Fe/H] = $-2.20 \pm 0.07$, and
[Eu/Fe] = $+2.45 \pm 0.08$.
Lines of lighter elements are marked in black font, and
lines of \rpro\ elements are marked in red font.
 }
\end{figure*}

We observed \jtwo\ 
on 2021 December 05
with two short (300~s and 500~s) integrations 
using the
Magellan Inamori Kyocera Echelle (MIKE; \citealt{bernstein03})
spectrograph on the Landon Clay (Magellan~II) 6.5~m Telescope
at Las Campanas Observatory, Chile.
This spectrum was obtained using the 0\farcs7 slit,
resulting in a spectral resolving power of 
$R \equiv \lambda/\Delta\lambda = 41,000$ at wavelengths
$< 5000$~\AA\ and $R = 35,000$ at wavelengths $> 5000$~\AA.~
The high signal-to-noise (S/N) ratio of this spectrum, 
$\approx$~80 at 3950~\AA, enabled us to immediately
recognize the unprecedented situation, shown in 
Figure~\ref{fig:specplot},
wherein the Eu~\textsc{ii} line at 3819.67~\AA\
appeared stronger than the Fe~\textsc{i} line at 3820.42~\AA.~
Unfortunately, poor weather prevented us from obtaining
additional followup during this observing run.

We reobserved \jtwo\ with MIKE on
2022 July 22 (4.61~hr) and
2022 August 18 (4.33~hr).
Both sets of observations were made with the 0\farcs35 slit, 
binned $1\times1$, yielding
$R = 60,000$ in the blue and
$R = 56,000$ in the red.
We used a non-standard blue grating setting to observe
orders 109 to 75 (3260--4780~\AA) on both nights.
On the first night, we also used a non-standard red grating setting to observe
orders 72 to 48 (4750--7250~\AA).~
On the second night we used the standard red grating setting to observe
orders 70 to 36 (4870--9650~\AA).
The orders redward of $\approx$~8000~\AA\ are 
marked by fringing, which we have not attempted to correct.
We used the CarPy MIKE reduction pipeline \citep{kelson00,kelson03}
to perform standard data reduction steps,
including overscan subtraction, pixel-to-pixel flat field division,
image coaddition, cosmic-ray removal, sky and scattered light subtraction,
rectification of the tilted slit profiles, spectrum extraction, and
wavelength calibration based on ThAr lamp spectra
collected immediately before or after our 
science exposures.
We perform the 
spectrum coaddition, order stitching, and continuum normalization using
routines in the 
Image Reduction and Analysis Facility (IRAF) ``echelle'' package
\citep{tody93}.
The S/N per pixel in the coadded spectrum ranges from
70 at 3500~\AA, 
190 at 3950~\AA,
290 at 4550~\AA, to
$>$300 at wavelengths longer than 5000~\AA.~

\subsection{Radial Velocity}
\label{sec:rv}

We compute the radial velocity, $v_{r}$, of each of the
three observations of \jtwo\ using
the echelle order containing the
Mg~\textsc{i} \textit{b} triplet ($\approx$~5150--5250~\AA).~
We use the IRAF ``fxcor'' task to
cross-correlate this order
against a metal-poor stellar template
observed using MIKE and shifted to rest velocity
\citep{roederer14c}.
The individual $v_{r}$ measurements are listed
in a footnote to Table~\ref{tab:data}.
All agree to within their mutual uncertainties,
and no variations are evident from the 
three observations spanning 8.5~mo.
Their mean velocity is 
3.1~\kmsec\ different from the $v_{r}$ measured by RAVE and
1.6~\kmsec\ different from the $v_{r}$ presented in Gaia's
third data release (DR3; \citealt{gaia22dr3rv}).~
We observed one radial-velocity standard star on each night
using the same MIKE setup, and 
the $v_{r}$ for each of those stars 
agrees with the corresponding Gaia velocity to within 
$\approx$~0.3~\kmsec.
These measurements could signal that
\jtwo\ is in a binary or multiple star system,
and we intend to perform
periodic radial velocity monitoring of \jtwo.

\subsection{The Rich Spectrum of J2213--5137}
\label{sec:inspect}

Figure~\ref{fig:specplot} shows three regions of the 
spectrum of \jtwo\ that include numerous lines
of heavy elements.
Two other stars with similar \teff\ and \logg\ values
are shown for comparison.
\hdone\ is only mildly enhanced in \rpro\ elements
([Eu/Fe] = $+0.29 \pm 0.16$; \citealt{roederer14c}),
while \hdtwo\ is highly enhanced in \rpro\ elements
([Eu/Fe] = $+ 1.32 \pm 0.08$; \citealt{roederer22a}).
\jtwo\ is more metal-poor
([Fe/H] = $-2.20 \pm 0.07$; Section~\ref{sec:modelatm})
than either \hdone\
([Fe/H] = $-1.80 \pm 0.08$)
or \hdtwo\
([Fe/H] = $-1.46 \pm 0.10$).
This fact is demonstrated in Figure~\ref{fig:specplot}
by the relative strengths of
lines of iron-group elements.
In contrast,
the lines of \rpro\ elements are substantially stronger
in the spectrum of \jtwo\ 
than in the spectrum of either comparison star.

As we will show in Section~\ref{sec:results},
\jtwo\ exhibits an \rpro\ enhancement ([Eu/Fe]) that is
about 13 times higher than \hdtwo\ and
about 280 times higher than the Sun.
Similarly,
\jtwo\ exhibits an \rpro\ enrichment ([Eu/H]) that is
about 2.4 times higher than \hdtwo\ and
about 1.8 times higher than the Sun.
In other words, per unit mass there are
nearly twice as many atoms of \rpro\ elements
in \jtwo\ than in the Sun, 
despite the overall factor of 160 difference in metallicity.
This contrast is evident among the 
lanthanide and actinide elements,
which exhibit numerous strong lines in the spectral regions
shown in Figure~\ref{fig:specplot}.
The spectrum of \jtwo\ presents a remarkable opportunity to
study the \rpro\ abundance pattern.

\section{Analysis}
\label{sec:analysis}

\subsection{Model Atmosphere Parameters}
\label{sec:modelatm}

We measure equivalent widths (EWs) of Fe~\textsc{i} and \textsc{ii} lines
using a semi-automated routine 
that fits line profiles in continuum-normalized spectra \citep{roederer14c}.
This routine uses an iterative clipping procedure to identify the local 
continuum of an $\approx$7~\AA\ region around the line of interest,
which can subsequently be modified by the user if necessary.
It then fits a Voigt profile, which defaults to a Gaussian profile for
the vast majority of lines, to the line of interest, and it 
integrates the EW under this fit.

Table~\ref{tab:lines} lists the 
line wavelengths ($\lambda$),
excitation potential (E.P.)\ of the lower level,
\loggf\ values, and
references for the \loggf\ values.
Table~\ref{tab:lines} also lists any 
hyperfine splitting (HFS) and isotope shifts (IS)
used to compute the line component patterns,
equivalent widths (EW),
abundances computed assuming local thermodynamic equilibrium (LTE),
upper limit (U.L.)\ flags,
and
non-LTE (NLTE) corrections.
The abundances derived from these lines,
along with broadband photometry and
astrometry from the 
Gaia mission \citep{gaia16main,gaia22dr3}
and other public resources,
help determine the model atmosphere parameters.

We use the LTE line analysis code MOOG
\citep{sneden73,sobeck11}
to derive abundances from these EWs and 
a model atmosphere.
We interpolate a model atmosphere from the
$\alpha$-enhanced grid of one-dimensional
ATLAS9 models \citep{castelli04}
using a code provided by A.\ McWilliam (2009, private communication).

We calculate the effective temperature, \teff,
using the color-metallicity-\teff\ calibrations 
provided by \citet{mucciarelli21} 
for Gaia and 2MASS broadband photometry.
We adopt an initial metallicity estimate of [Fe/H] $= -2.0$.
Six color combinations are constructed among the
Gaia $G$, $G_{BP}$, $G_{RP}$, and 2MASS $K_{s}$ magnitudes.
We deredden these colors using the 
$E(B-V)$ color excess inferred from the dust maps of \citet{schlafly11}
and the extinction coefficients presented by \citet{gaia18c}.
The amount of interstellar absorption at the stronger
component of the Na~\textsc{i} \textit{D} doublet
does not exceed 45~m\AA.
This translates \citep{roederer18b}
to a low $E(B-V)$, $< 0.01$~mag or so,
supporting the low extinction value inferred from the dust maps
(Table~\ref{tab:data}).

We estimate the statistical uncertainty in \teff\ by
resampling each of these input parameters and 
recalculating \teff\ $10^{4}$ times.
Each parameter is resampled from a Gaussian distribution
with a mean equal to the parameter's value and
a standard deviation equal to the parameter's 1$\sigma$ uncertainty.
We calculate the weighted mean of the six
predicted \teff\ values, 
5509~K, and we adopt it
as the final \teff\ for \jtwo.
We estimate the systematic uncertainty in \teff\ by
performing analogous calculations using the 
color-metallicity-\teff\ calibrations presented by
\citet{alonso99b}, \citet{ramirez05b}, and \citet{casagrande10}.
These studies calibrated several sets of colors among
$B$ and $V$ Johnson photometry and $J$, $H$, and $K_{s}$ 
(or $K_{\rm TCS}$) photometry.
These alternative scales predict \teff\ = 
$5467 \pm 108$~K, 
$5417 \pm 58$~K, and
$5612 \pm 67$~K, respectively.
We adopt the standard deviation of these four
values as an approximate lower limit on
the systematic uncertainty in \teff.
Table~\ref{tab:data} lists our adopted \teff\ value for \jtwo,
$5509 \pm 43$ (stat.)\ $\pm 72$ (sys.)~K.

We calculate the log of the surface gravity, \logg, 
using the fundamental relations expressed in
equation~1 of \citet{roederer18c}.
The relevant quantities for \jtwo\ are listed in Table~\ref{tab:data},
and we assume a mass of $0.80 \pm 0.08$~\msun.
We resample each of the input parameters $10^{4}$ times.
We adopt the median value of this distribution as \logg, 
and we adopt the standard deviation of these values as its uncertainty:\
$2.28 \pm 0.06$.
The \teff\ and \logg\ of \jtwo\ imply that it is
a field-star equivalent of the red horizontal branch
found in globular clusters.

We interpolate a model atmosphere for these \teff\ and \logg\ values,
and we use MOOG to derive the abundance
from each Fe~\textsc{i} and \textsc{ii} line.
We adopt the microturbulent velocity parameter, \vt, 
that produces no correlation between the strength of Fe~\textsc{i} lines
and the abundances derived from them.
We iterate the calculations for \teff, \logg, and \vt\
with updated [Fe/H] values.
Departures from LTE (i.e., NLTE) 
affect the Fe abundance derived from Fe~\textsc{i} lines
(e.g., \citealt{thevenin99}).
We use the INSPECT database\footnote{%
\url{http://www.inspect-stars.com}}
\citep{bergemann12,lind12} to
compute NLTE corrections for 15 Fe~\textsc{i} lines
included in that database.
These corrections,
listed in Table~\ref{tab:lines},
average $+0.19 \pm 0.01$~dex.
This value is in excellent agreement with the
empirical correction predicted by equation~1 of
\citet{ezzeddine17}, $+0.18$~dex.
The NLTE-corrected [Fe/H] ratio derived from Fe~\textsc{i} lines
is $-2.20 \pm 0.12$.
The [Fe/H] ratio derived from Fe~\textsc{ii} lines, 
which should be minimally affected by NLTE,
is $-2.25 \pm 0.06$.
The two values are in excellent agreement, and we adopt 
[Fe/H] $= -2.20 \pm 0.12$ as the metallicity of \jtwo.

\startlongtable
\begin{deluxetable*}{ccccccccc}
\tablecaption{Line Atomic Data and Derived Abundances
\label{tab:lines}}
\tablewidth{0pt}
\tabletypesize{\small} 
\tablehead{
\colhead{Species} &
\colhead{$\lambda$} &
\colhead{E.P.} &
\colhead{\loggf} &
\colhead{\loggf} &
\colhead{EW} &
\colhead{U.L.} &
\colhead{$\log\varepsilon$} &
\colhead{NLTE} \\
\colhead{} &
\colhead{(\AA)} &
\colhead{(eV)} &
\colhead{} &
\colhead{ref.} &
\colhead{(m\AA)} &
\colhead{flag} &
\colhead{[LTE]} &
\colhead{corr.} 
}
\startdata
  Li~\textsc{i} &   6707.80 &     0.00 &     0.17 &  1  &         &$<$& 0.70 &         \\
   O~\textsc{i} &   7771.94 &     9.15 &     0.35 &  2  &    20.5 &   & 7.32 & $-$0.14 \\
   O~\textsc{i} &   7774.17 &     9.15 &     0.20 &  2  &    15.9 &   & 7.32 & $-$0.14 \\
   O~\textsc{i} &   7775.39 &     9.15 &    -0.02 &  2  &    13.2 &   & 7.44 & $-$0.14 \\
  Na~\textsc{i} &   5889.95 &     0.00 &     0.11 &  3  &   187.8 &   & 4.62 & $-$0.68 \\
  Na~\textsc{i} &   5895.92 &     0.00 &    -0.19 &  3  &   168.6 &   & 4.65 & $-$0.63 \\
\enddata      
\tablecomments{The NLTE-corrected abundances may be obtained by adding the NLTE corrections
to the LTE $\log\varepsilon$ abundances.
The complete version of Table~\ref{tab:lines} is available 
in machine-readable form in the online edition of the journal.
A portion is shown here to illustrate its form and content.}
\tablereferences{%
1:\ \citet{smith98}, using HFS from \citet{kurucz11};
2:\ \citet{magg22};
3:\ \citet{kramida21} (NIST ASD);
4:\ \citet{pehlivanrhodin17};
5:\ \citet{kramida21}, using HFS from VALD3 \citep{piskunov95,pakhomov19};
6:\ \citet{denhartog21ca};
7:\ \citet{lawler89}, using HFS from \citet{kurucz11};
8:\ \citet{lawler13};
9:\ \citet{wood13};
10:\ \citet{lawler14}, including HFS;
11:\ \citet{wood14v}, including HFS;
12:\ \citet{sobeck07};
13:\ \citet{lawler17};
14:\ \citet{denhartog11}, including HFS;
15:\ \citet{obrian91};
16:\ \citet{belmonte17};
17:\ \citet{denhartog14};
18:\ \citet{ruffoni14};
19:\ \citet{melendez09fe};
20:\ \citet{denhartog19};
21:\ \citet{lawler15}, including HFS;
22:\ \citet{wood14ni};
23:\ \citet{kramida21}, using HFS from \citet{kurucz11};
24:\ \citet{roederer12b};
25:\ \citet{li99};
26:\ \citet{morton00};
27:\ \citet{biemont11};
28:\ \citet{ljung06};
29:\ \citet{nilsson10};
30:\ \citet{wickliffe94};
31:\ \citet{duquette85};
32:\ \citet{biemont82};
33:\ \citet{hansen12}, including HFS and IS;
34:\ \citet{kramida21}, using HFS/IS from \citet{mcwilliam98} when available;
35:\ \citet{lawler01la}, using HFS from \citet{ivans06} when available;
36:\ \citet{lawler09};
37:\ \citet{li07}, using HFS from \citet{sneden09};
38:\ \citet{ivarsson01}, using HFS from \citet{sneden09};
39:\ \citet{denhartog03}, using HFS and IS from \citet{roederer08a} when 
        available;
40:\ \citet{lawler06}, using HFS and IS from \citet{roederer08a} when
        available;
41:\ \citet{lawler01eu}, using HFS and IS from \citet{ivans06};
42:\ \citet{denhartog06};
43:\ \citet{lawler01tb}, using HFS from \citet{lawler01tbhfs};
44:\ \citet{wickliffe00};
45:\ \citet{lawler04}, including HFS;
46:\ \citet{lawler08};
47:\ \citet{wickliffe97tm};
48:\ \citet{sneden09}, including HFS and IS;
49:\ \citet{lawler09}, including HFS;
50:\ \citet{lawler07};
51:\ \citet{quinet06};
52:\ \citet{xu07}, using HFS and IS from \citet{cowan05};
53:\ \citet{denhartog05}, including HFS and IS;
54:\ \citet{biemont00}, including HFS and IS from \citet{roederer12d};
55:\ \citet{nilsson02th};
56:\ \citet{nilsson02u}.
}
\end{deluxetable*}

\subsection{Abundance Analysis}
\label{sec:abundanalysis}

We measure EWs for
O~\textsc{i},
Na~\textsc{i},
Mg~\textsc{i},
Si~\textsc{i},
K~\textsc{i},
Ca~\textsc{i},
Ti~\textsc{i} and \textsc{ii},
Cr~\textsc{i} and \textsc{ii},
Ni~\textsc{i}, and 
Zn~\textsc{i}
using the same methods described in Section~\ref{sec:modelatm}.
We derive abundances from these EWs using the MOOG ``abfind'' driver,
the model atmosphere derived in Section~\ref{sec:modelatm}, and 
the atomic data presented in Table~\ref{tab:lines}.
We derive abundances or upper limits
for all other species using the MOOG ``synth'' driver.
We generate line lists for the synthetic spectrum matching
using the LINEMAKE code\footnote{%
\url{https://github.com/vmplacco/linemake}}
\citep{placco21linemake}.
We include multiple isotopes in our syntheses for the elements
Li ($^{6}$Li/$^{7}$Li = 0; \citealt{lind13}), 
C ($^{12}$C/$^{13}$C = 5, typical for an evolved star;
 \citealt{gratton00}), 
N ($^{15}$N/$^{14}$N = 0; \citealt{meija16}), 
Cu ($^{63}$Cu/$^{65}$Cu = 2.24, the Solar ratio; \citealt{meija16}), 
Ag, Ba, Nd, Sm, Eu, Yb, Ir, Pt, and Pb.
We adopt the \rpro\ isotopic ratios for Ag through Pb
\citep{sneden08}.
We apply NLTE corrections for lines of
O~\textsc{i} \citep{amarsi15},
Na~\textsc{i} \citep{lind11},
Al~\textsc{i} \citep{nordlander17al,roederer21},
K~\textsc{i} \citep{takeda02},
Ti~\textsc{i} \citep{bergemann11},
Cu~\textsc{i} \citep{roederer18a,shi18},
and
Pb~\textsc{i} \citep{mashonkina12,roederer20}.

Whenever possible, we derive abundances from features
with $\lambda < 3640$~\AA\ or $\lambda > 4400$~\AA,
where the density of lanthanide lines is lower and
absorption from the convergence of Balmer-series transitions
does not complicate continuum identification or modeling.
516 lines in our spectrum of \jtwo\
yield reliable estimates of the
abundances (and upper limits) of 
47 (53) elements, including
29 (33) elements produced by \rpro\ nucleosynthesis.
Table~\ref{tab:abund} lists the final abundances,
including NLTE corrections.

\begin{deluxetable}{cccccccc}
\tablecaption{Derived Abundances
\label{tab:abund}}
\tablewidth{0pt}
\tabletypesize{\scriptsize}
\tablehead{
\colhead{Species} &
\colhead{$Z$} &
\colhead{$\log\varepsilon_{\odot}$} &
\colhead{$\log\varepsilon$(X)} &
\colhead{[X/Fe]\tablenotemark{a}} &
\colhead{$\sigma$($\log\varepsilon$(X))} &
\colhead{$\sigma$([X/Fe])} &
\colhead{$N$} 
}
\startdata
Li~\textsc{i}  &  3 &\nodata&$<$0.70& \nodata  & \nodata& \nodata  & 1   \\
C~(CH)         &  6 & 8.43 &  6.61  &  $+$0.38 &   0.30 &   0.30   & 1   \\
N~(NH)         &  7 & 7.83 & $<$6.30& $<+$0.67 & \nodata& \nodata  & 1   \\
O~\textsc{i}   &  8 & 8.69 &  7.22  &  $+$0.73 &   0.15 &   0.26   & 3   \\
Na~\textsc{i}  & 11 & 6.24 &  3.98  &  $-$0.06 &   0.17 &   0.07   & 2   \\
Mg~\textsc{i}  & 12 & 7.60 &  5.89  &  $+$0.49 &   0.11 &   0.05   & 4   \\
Al~\textsc{i}  & 13 & 6.45 &  3.87  &  $-$0.38 &   0.17 &   0.08   & 1   \\
Si~\textsc{i}  & 14 & 7.51 &  5.63  &  $+$0.32 &   0.20 &   0.22   & 4   \\
Si~\textsc{ii} & 14 & 7.51 &  5.70  &  $+$0.39 &   0.17 &   0.19   & 2   \\
K~\textsc{i}   & 19 & 5.03 &  2.98  &  $+$0.15 &   0.11 &   0.04   & 2   \\
Ca~\textsc{i}  & 20 & 6.34 &  4.39  &  $+$0.25 &   0.09 &   0.05   & 17  \\
Sc~\textsc{ii} & 21 & 3.15 &  1.08  &  $+$0.13 &   0.08 &   0.07   & 3   \\
Ti~\textsc{i}  & 22 & 4.95 &  3.13  &  $+$0.38 &   0.15 &   0.06   & 10  \\
Ti~\textsc{ii} & 22 & 4.95 &  3.02  &  $+$0.27 &   0.07 &   0.06   & 4   \\
V~\textsc{i}   & 23 & 3.93 &  1.82  &  $+$0.09 &   0.17 &   0.08   & 2   \\
V~\textsc{ii}  & 23 & 3.93 &  1.85  &  $+$0.12 &   0.14 &   0.13   & 3   \\
Cr~\textsc{i}  & 24 & 5.64 &  3.18  &  $-$0.26 &   0.14 &   0.06   & 4   \\
Cr~\textsc{ii} & 24 & 5.64 &  3.46  &  $+$0.02 &   0.10 &   0.10   & 2   \\
Mn~\textsc{i}  & 25 & 5.43 &  2.88  &  $-$0.35 &   0.20 &   0.14   & 3   \\
Mn~\textsc{ii} & 25 & 5.43 &  2.86  &  $-$0.37 &   0.12 &   0.11   & 1   \\
Fe~\textsc{i}  & 26 & 7.50 &  5.30  &  $-$2.20 &   0.12 &   0.12   & 99  \\
Fe~\textsc{ii} & 26 & 7.50 &  5.25  &  $-$2.25 &   0.06 &   0.06   & 11  \\
Co~\textsc{i}  & 27 & 4.99 &  2.66  &  $-$0.13 &   0.22 &   0.12   & 7   \\
Ni~\textsc{i}  & 28 & 6.22 &  3.90  &  $-$0.12 &   0.11 &   0.06   & 12  \\
Cu~\textsc{i}  & 29 & 4.19 &  1.22  &  $-$0.77 &   0.22 &   0.12   & 1   \\
Zn~\textsc{i}  & 30 & 4.56 &  2.52  &  $+$0.16 &   0.10 &   0.07   & 2   \\
Ge~\textsc{i}  & 32 & 3.65 & $<$2.40& $<+$0.95 & \nodata& \nodata  & 1   \\
Rb~\textsc{i}  & 37 & 2.52 & $<$2.20& $<+$1.88 & \nodata& \nodata  & 1   \\
Sr~\textsc{ii} & 38 & 2.87 &  1.79  &  $+$1.12 &   0.18 &   0.18   & 2   \\
Y~\textsc{ii}  & 39 & 2.21 &  1.14  &  $+$1.13 &   0.08 &   0.07   & 23  \\
Zr~\textsc{ii} & 40 & 2.58 &  1.87  &  $+$1.49 &   0.09 &   0.08   & 9   \\
Nb~\textsc{ii} & 41 & 1.46 &  0.90  &  $+$1.64 &   0.18 &   0.18   & 3   \\
Mo~\textsc{i}  & 42 & 1.88 &  1.34  &  $+$1.66 &   0.27 &   0.20   & 2   \\
Ru~\textsc{i}  & 44 & 1.75 &  1.59  &  $+$2.04 &   0.19 &   0.08   & 4   \\
Rh~\textsc{i}  & 45 & 0.91 &  0.99  &  $+$2.28 &   0.23 &   0.13   & 3   \\
Pd~\textsc{i}  & 46 & 1.57 &  1.15  &  $+$1.78 &   0.21 &   0.10   & 3   \\
Ag~\textsc{i}  & 47 & 0.94 &  0.50  &  $+$1.76 &   0.22 &   0.12   & 1   \\
Ba~\textsc{ii} & 56 & 2.18 &  1.70  &  $+$1.72 &   0.13 &   0.12   & 6   \\
La~\textsc{ii} & 57 & 1.10 &  0.87  &  $+$1.97 &   0.11 &   0.10   & 24  \\
Ce~\textsc{ii} & 58 & 1.58 &  1.07  &  $+$1.69 &   0.10 &   0.09   & 19  \\
Pr~\textsc{ii} & 59 & 0.72 &  0.47  &  $+$1.95 &   0.11 &   0.10   & 9   \\
Nd~\textsc{ii} & 60 & 1.42 &  1.15  &  $+$1.93 &   0.11 &   0.10   & 43  \\
Sm~\textsc{ii} & 62 & 0.96 &  0.93  &  $+$2.17 &   0.11 &   0.10   & 38  \\
Eu~\textsc{ii} & 63 & 0.52 &  0.77  &  $+$2.45 &   0.09 &   0.08   & 10  \\
Gd~\textsc{ii} & 64 & 1.07 &  1.12  &  $+$2.25 &   0.12 &   0.10   & 34  \\
Tb~\textsc{ii} & 65 & 0.30 &  0.29  &  $+$2.19 &   0.13 &   0.11   & 5   \\
Dy~\textsc{ii} & 66 & 1.10 &  1.24  &  $+$2.34 &   0.16 &   0.15   & 15  \\
Ho~\textsc{ii} & 67 & 0.48 &  0.46  &  $+$2.18 &   0.16 &   0.14   & 8   \\
Er~\textsc{ii} & 68 & 0.92 &  1.03  &  $+$2.31 &   0.13 &   0.12   & 11  \\
Tm~\textsc{ii} & 69 & 0.10 &  0.12  &  $+$2.22 &   0.12 &   0.11   & 9   \\
Yb~\textsc{ii} & 70 & 0.84 &  1.01  &  $+$2.37 &   0.20 &   0.19   & 1   \\
Lu~\textsc{ii} & 71 & 0.10 &  0.36  &  $+$2.46 &   0.08 &   0.07   & 4   \\
Hf~\textsc{ii} & 72 & 0.85 &  0.71  &  $+$2.06 &   0.12 &   0.11   & 5   \\
Os~\textsc{i}  & 76 & 1.40 &  1.60  &  $+$2.40 &   0.24 &   0.12   & 2   \\
Ir~\textsc{i}  & 77 & 1.38 &  1.45  &  $+$2.27 &   0.21 &   0.10   & 2   \\
Pt~\textsc{i}  & 78 & 1.62 & $<$1.95& $<+$2.53 & \nodata& \nodata  & 1   \\
Pb~\textsc{i}  & 80 & 2.04 & $<$1.50& $<+$1.66 & \nodata& \nodata  & 2   \\
Th~\textsc{ii} & 90 & 0.02 &  0.22  &  $+$2.40 &   0.13 &   0.12   & 8   \\
U~\textsc{ii}  & 92 &$-$0.54&$-$0.74&  $+$2.00 &   0.26 &   0.26   & 1   \\
\enddata      
\tablenotetext{a}{[Fe/H] is given instead of [X/Fe] for Fe.}
\tablecomments{%
The Solar abundances, listed in column~3,
are adopted from \citet{asplund09}.
The abundance of element X is defined
as the number of X atoms per 10$^{12}$ H atoms,
$\log\varepsilon$(X)~$\equiv \log_{10}(N_{\rm X}/N_{\rm H})+$12.0.
The C abundances have been corrected 
(by $+$0.71 dex) to the ``natal'' abundance
according to the stellar evolution corrections presented by
\citet{placco14c}.
A single C abundance is derived by synthesizing the
CH molecular features in the
spectral region from 4290--4330~\AA.~
A single N abundance is derived by synthesizing the
NH molecular features in the
spectral region from 3360--3370~\AA.~
NLTE corrections have been applied to the abundances derived from
O~\textsc{i}, Na~\textsc{i}, Al~\textsc{i}, K~\textsc{i}, 
Ti~\textsc{i}, and Fe~\textsc{i} lines.
}
\end{deluxetable}

We compute abundance uncertainties by resampling
the model atmosphere parameters and line data $10^{3}$ times,
following the method outlined in \citet{roederer18c}.
We conservatively assume that the systematic uncertainties
in \teff\ and \logg\ may be twice the size of the
lower limits adopted in Section~\ref{sec:modelatm}.
We also conservatively assume a minimum uncertainty, $\sigma_{\rm EW}$, of 
2~m\AA\ and 1\% in the EWs of lines,
as follows:\ $\sigma_{\rm EW} = \sqrt{2.0^{2} + (0.01\times{\rm EW})^{2}}$,
where all units are in m\AA.
This component accounts for sources of uncertainty that are relatively
uniform throughout the spectrum, such as flat-fielding errors, 
and the values adopted are based on our experience with the data.
Our calculations include a wavelength-dependent term that varies 
smoothly from blue to red to account for 
S/N and continuum placement,
$\sigma_{\rm EW} = 2\times10^{15} \lambda^{-4.2}$,
where $\lambda$ is measured in \AA\ and 
$\sigma_{\rm EW}$ is measured in m\AA.~
These values are also based on our experience with the data.
This component of the uncertainty is 
$\approx$~3~m\AA\ at 3400~\AA,
$\approx$~1.5~m\AA\ at 4000~\AA, and
$< 1$~m\AA\ at $\lambda > 4400$~\AA.
We add these two components of the uncertainty in quadrature.
We compute the unweighted mean abundance of each element, X, and the
[X/Fe] ratio for each resample.
The uncertainties listed in
Table~\ref{tab:abund}
reflect the inner 68$^{\rm th}$ percentiles of the distributions.

\begin{figure*}
\begin{center}
\includegraphics[angle=0,width=2.5in]{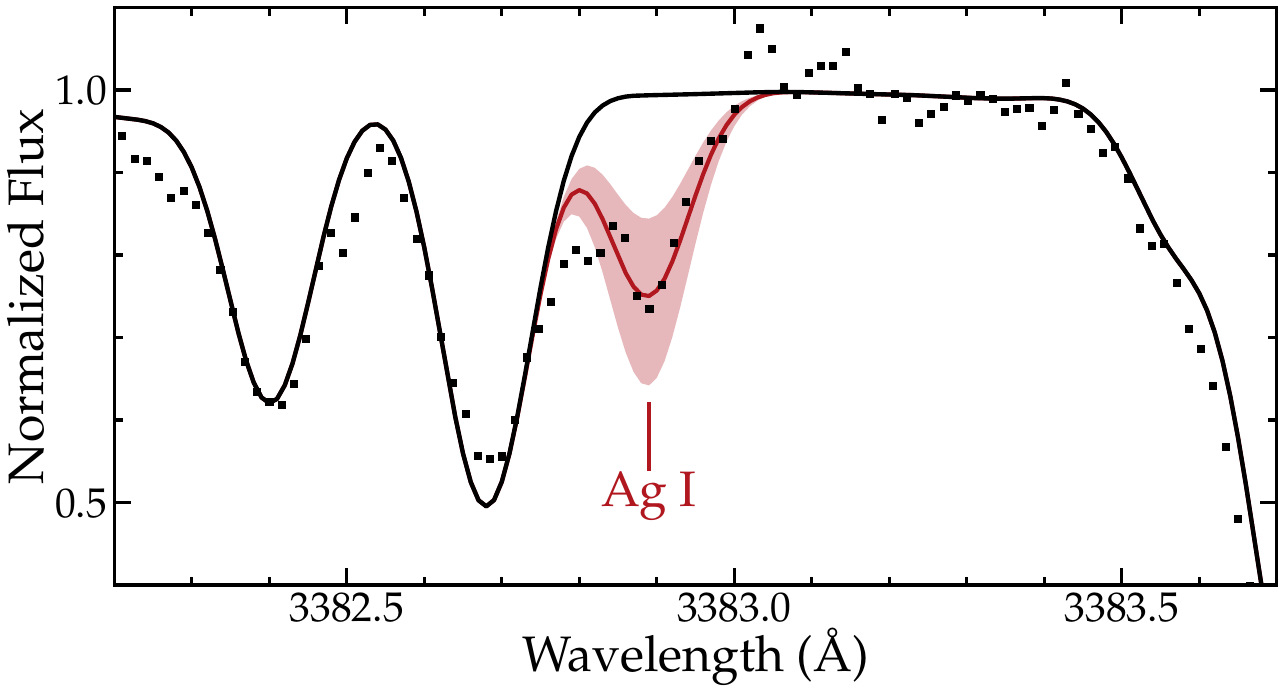}
\hspace*{0.1in}
\includegraphics[angle=0,width=2.5in]{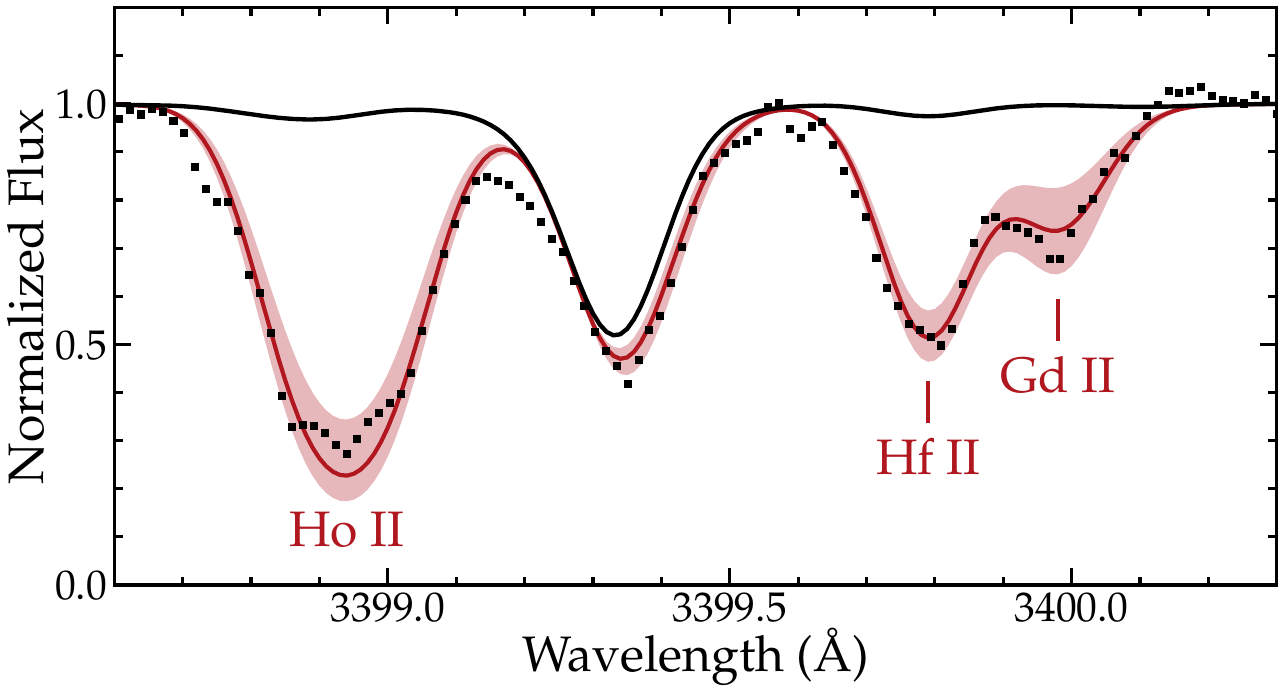} \\
\includegraphics[angle=0,width=2.5in]{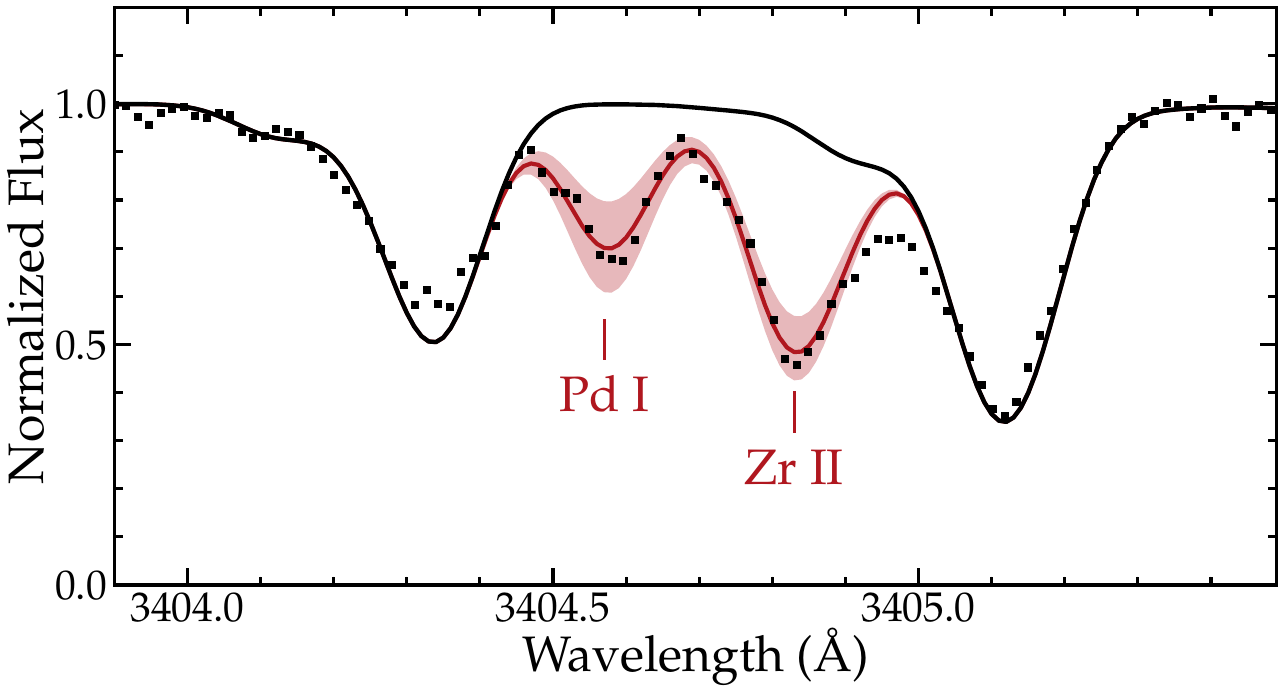}
\hspace*{0.1in}
\includegraphics[angle=0,width=2.5in]{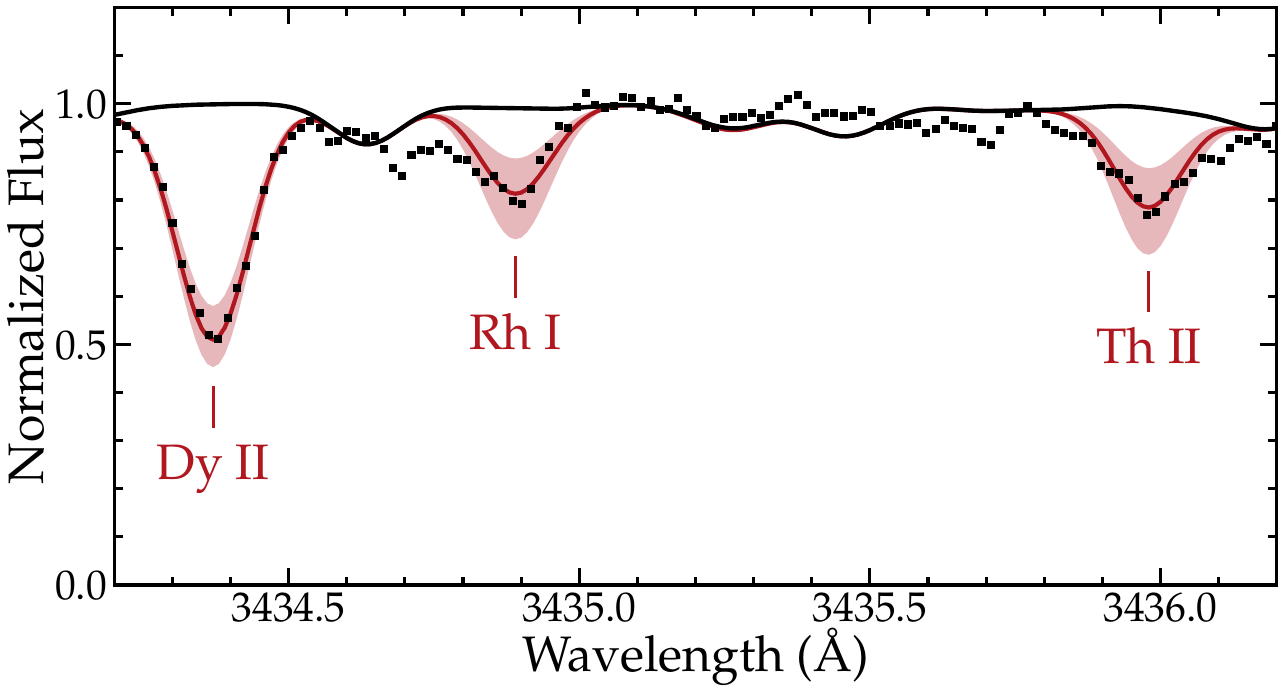} \\
\includegraphics[angle=0,width=2.5in]{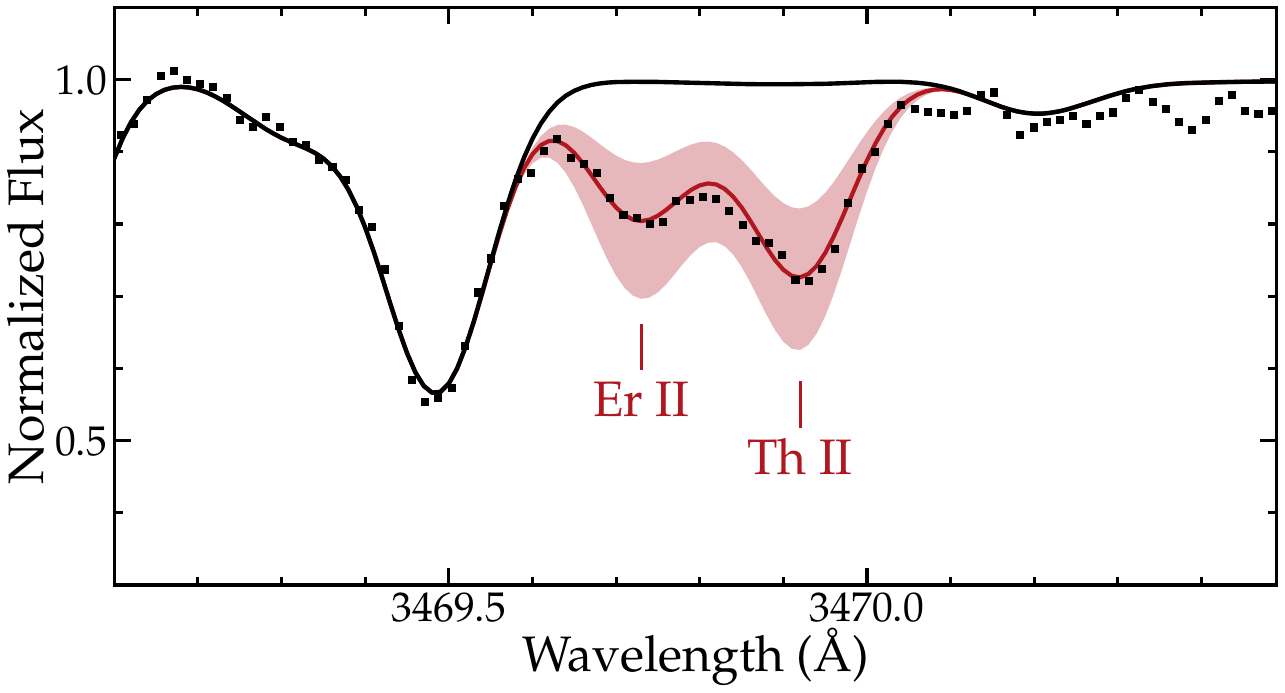}
\hspace*{0.1in}
\includegraphics[angle=0,width=2.5in]{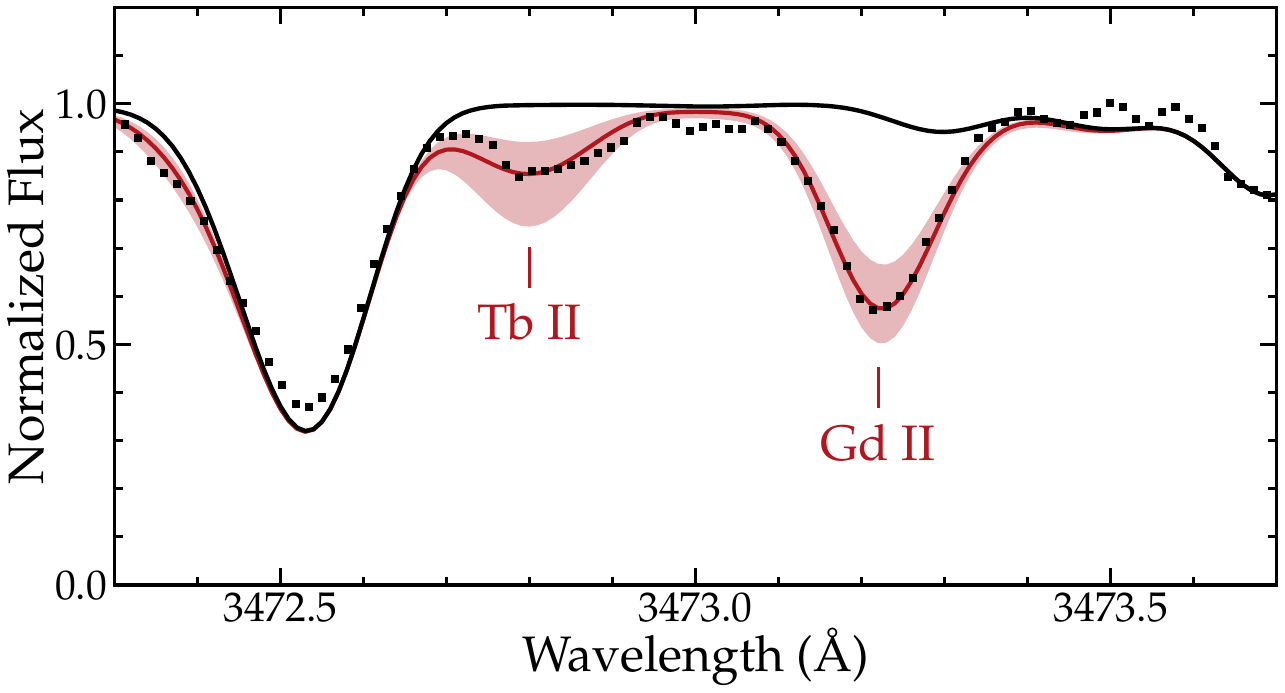} \\
\end{center}
\caption{
\label{fig:synth1}
Comparison of observed spectra (black dots)
and synthetic spectra for several lines of interest.
The red line marks the best-fit abundance 
to each absorption line, and the
light red band marks $\pm$~0.3~dex 
relative to the best-fit line.
The black lines are synthesized
without the element of interest.
 }
\end{figure*}

\begin{figure*}
\begin{center}
\includegraphics[angle=0,width=2.5in]{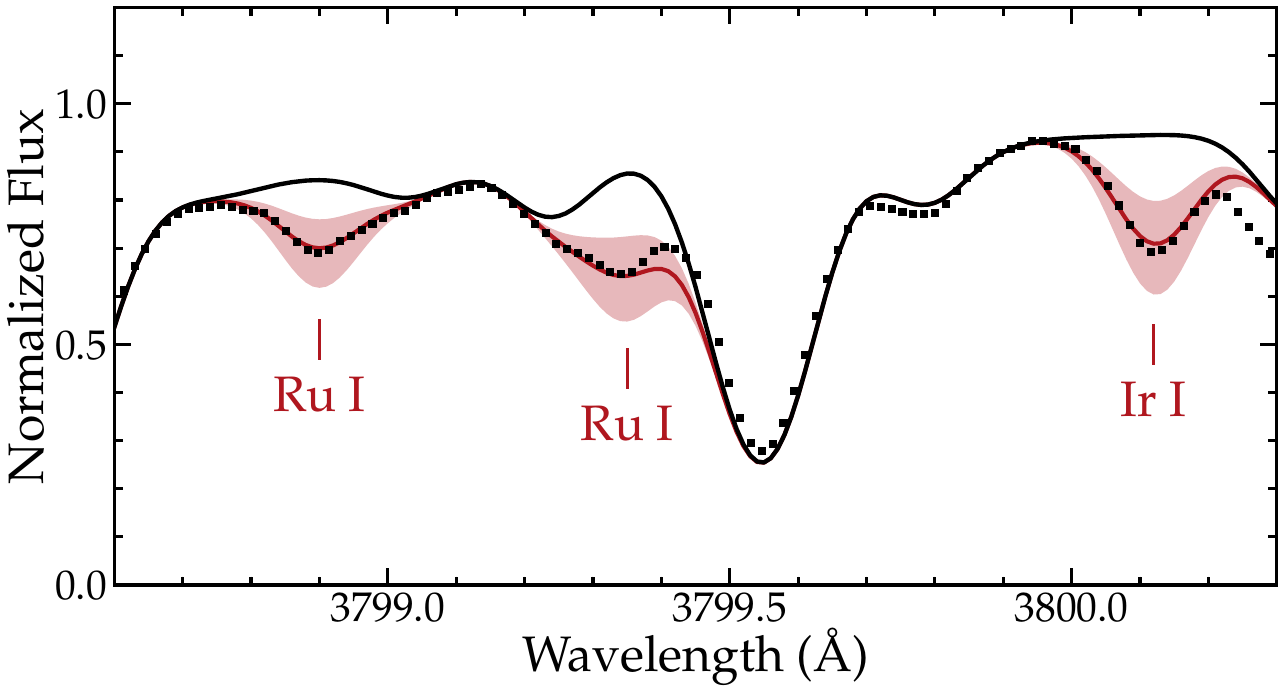}
\hspace*{0.1in}
\includegraphics[angle=0,width=2.5in]{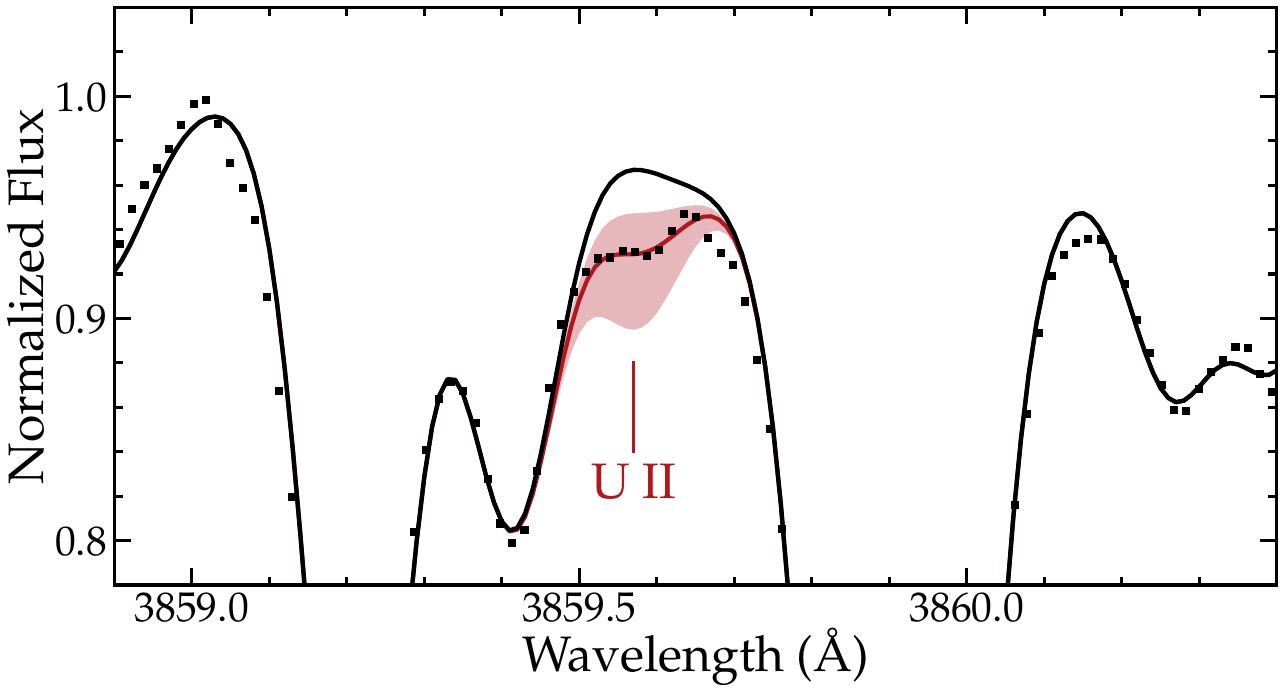} \\
\includegraphics[angle=0,width=2.5in]{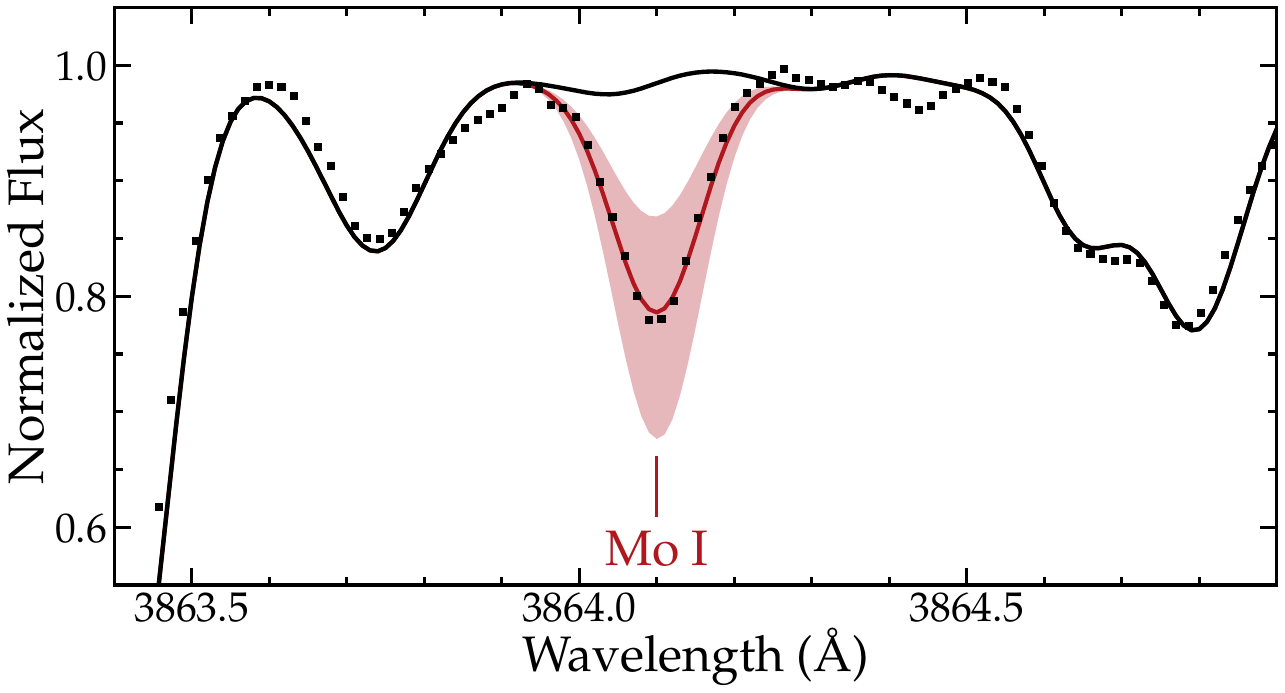}
\hspace*{0.1in}
\includegraphics[angle=0,width=2.5in]{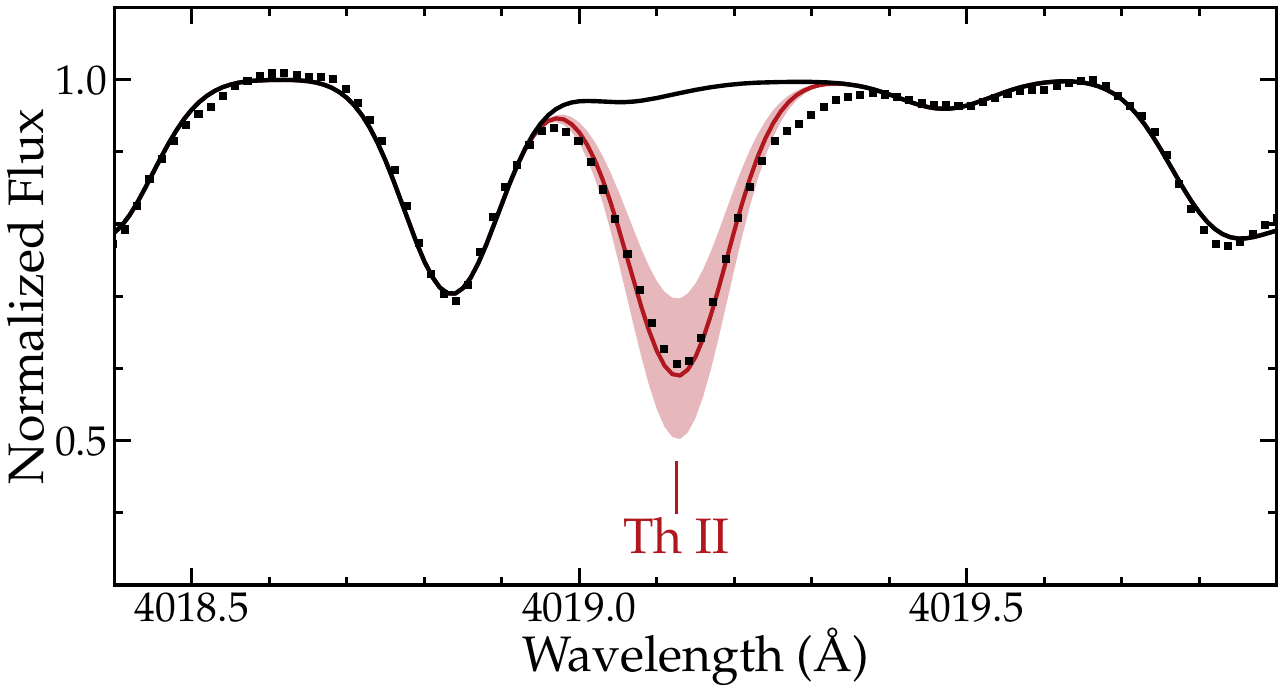} \\
\includegraphics[angle=0,width=2.5in]{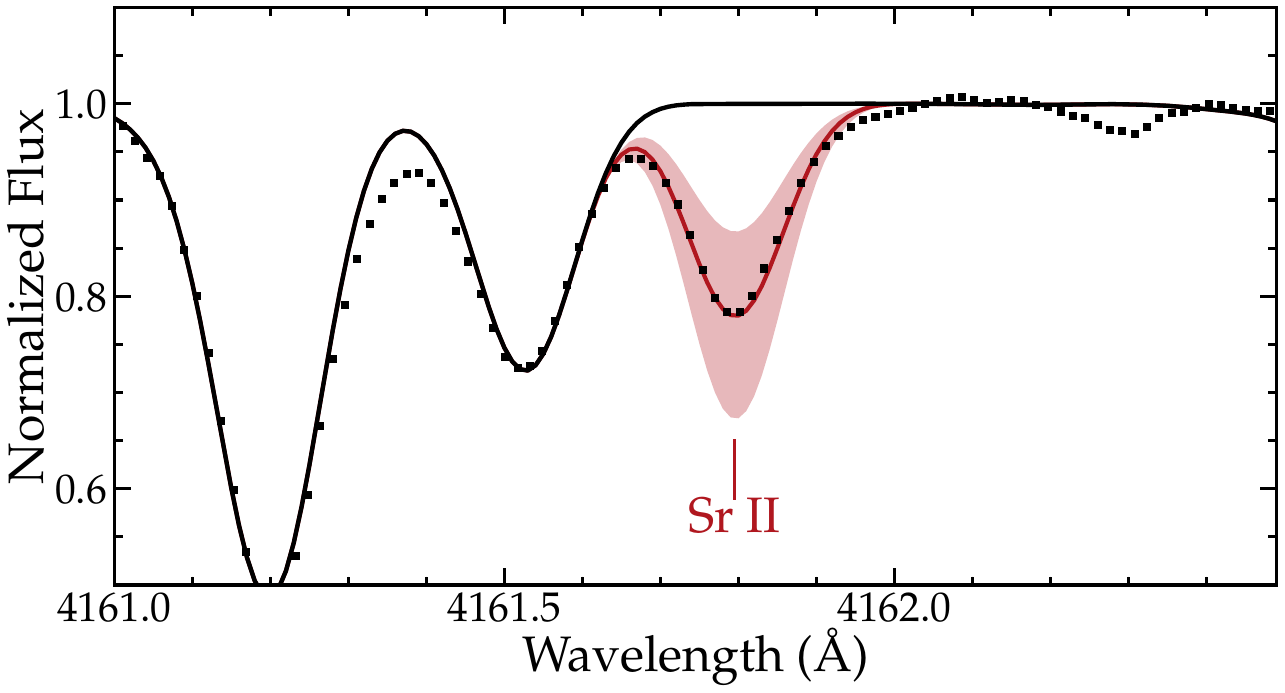}
\hspace*{0.1in}
\includegraphics[angle=0,width=2.5in]{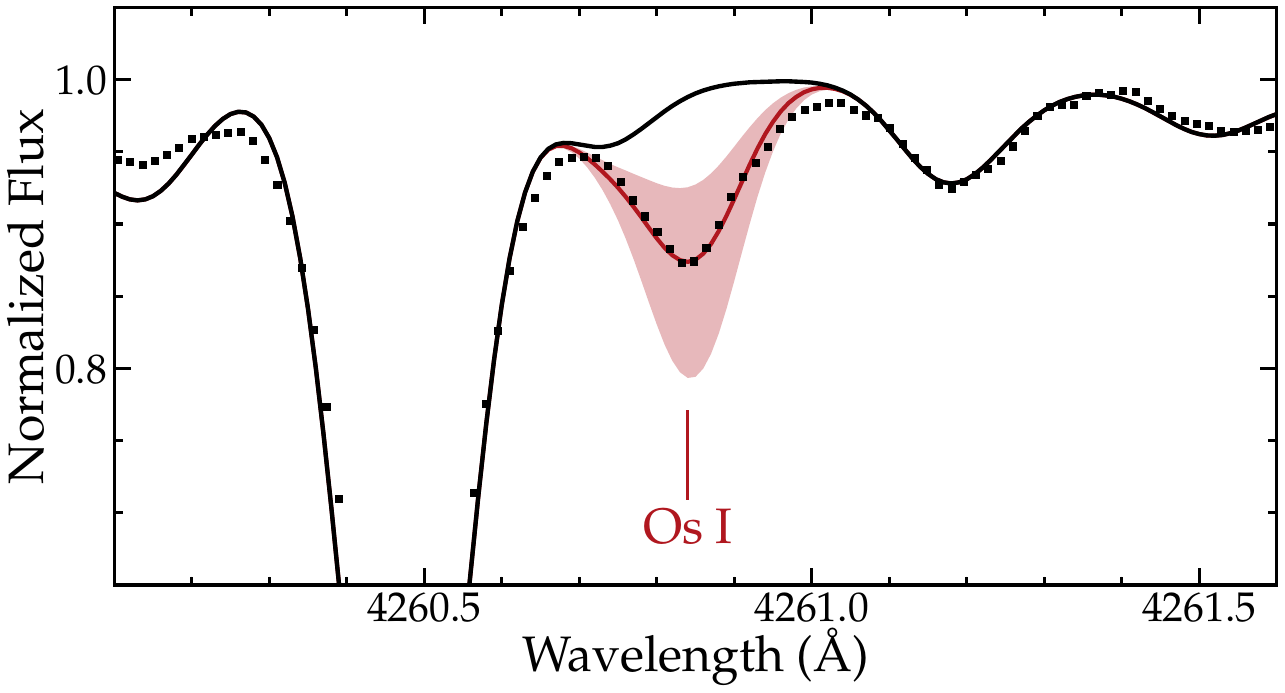} \\
\end{center}
\caption{
\label{fig:synth2}
Comparison of observed
and synthetic spectra for several lines of interest.
Symbols are the same as in Figure~\ref{fig:synth1}.
 }
\end{figure*}

\begin{figure*}
\begin{center}
\includegraphics[angle=0,width=2.5in]{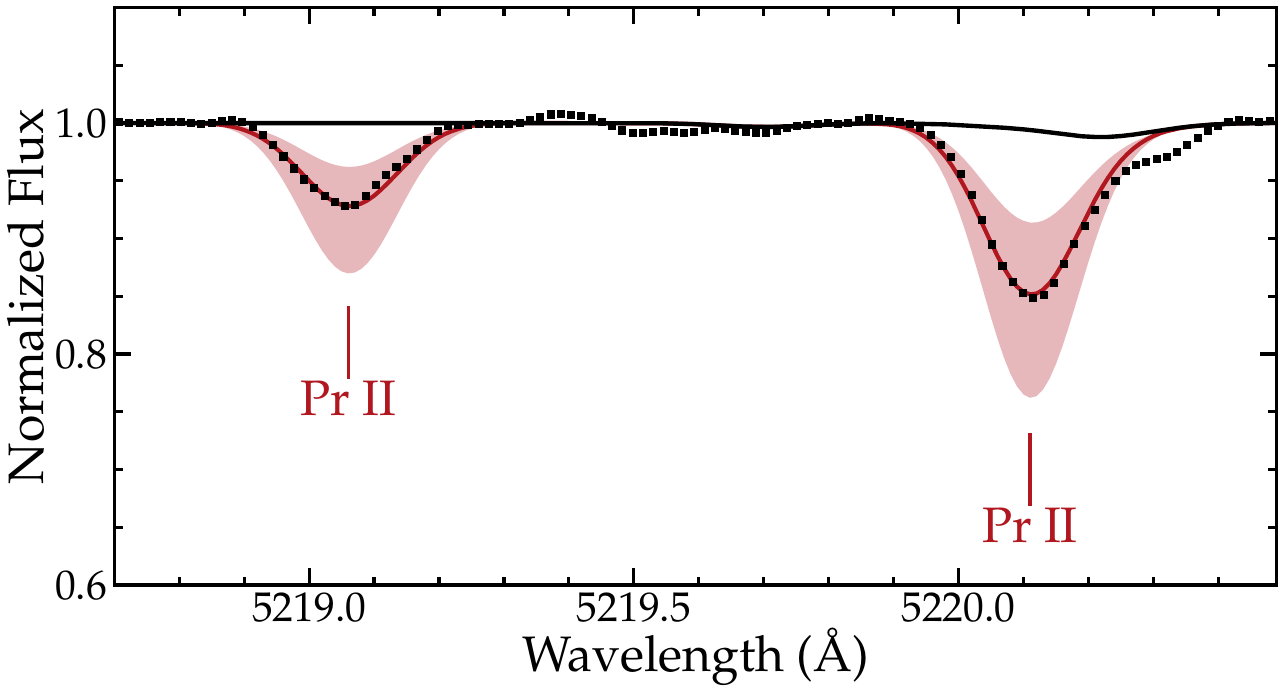}
\hspace*{0.1in}
\includegraphics[angle=0,width=2.5in]{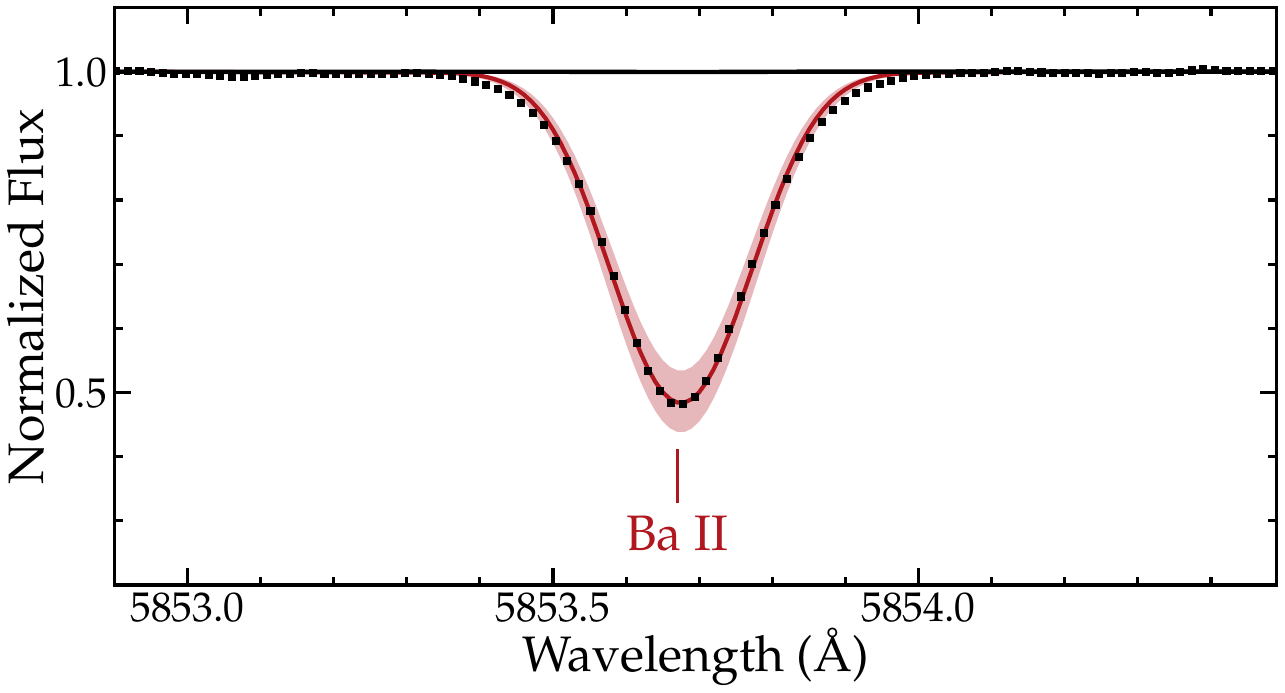} \\
\includegraphics[angle=0,width=2.5in]{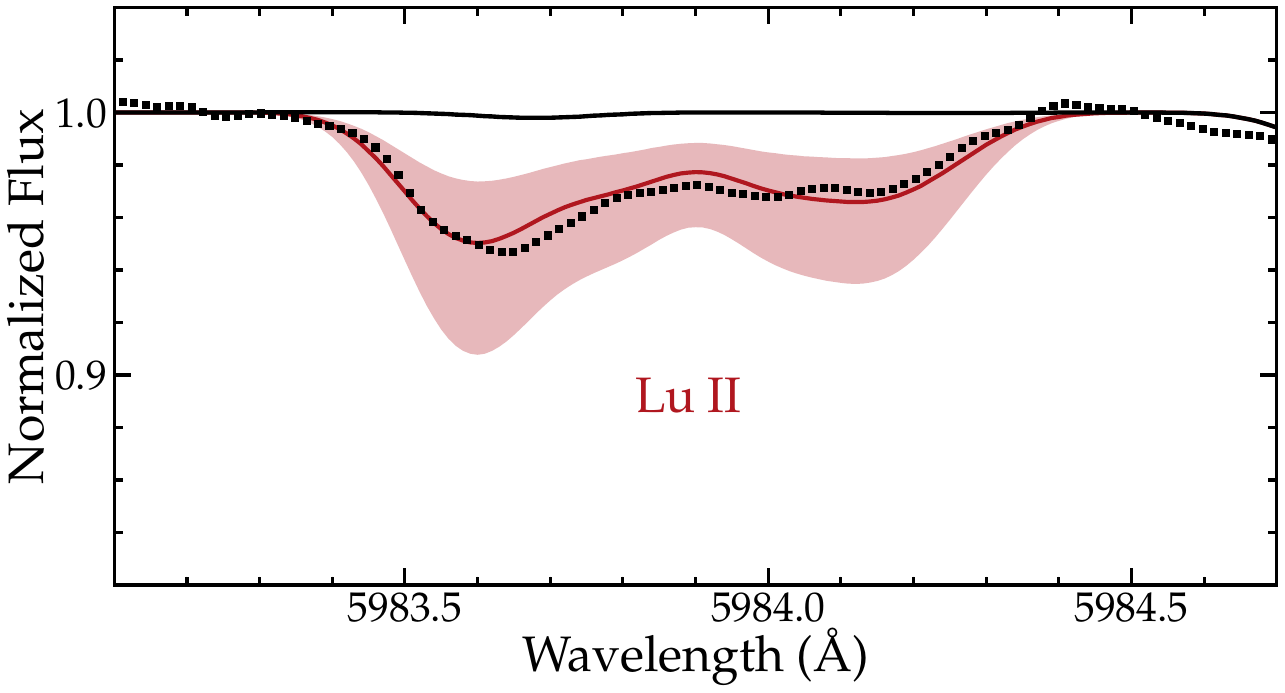}
\hspace*{0.1in}
\includegraphics[angle=0,width=2.5in]{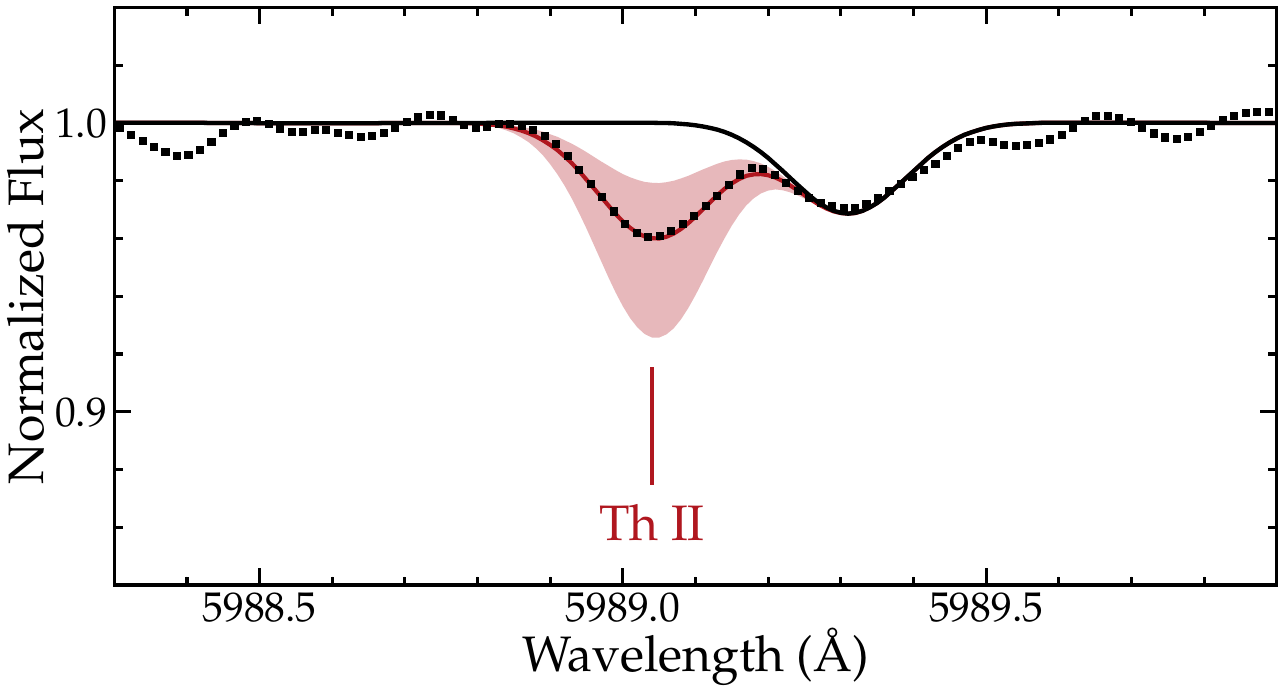} \\
\includegraphics[angle=0,width=2.5in]{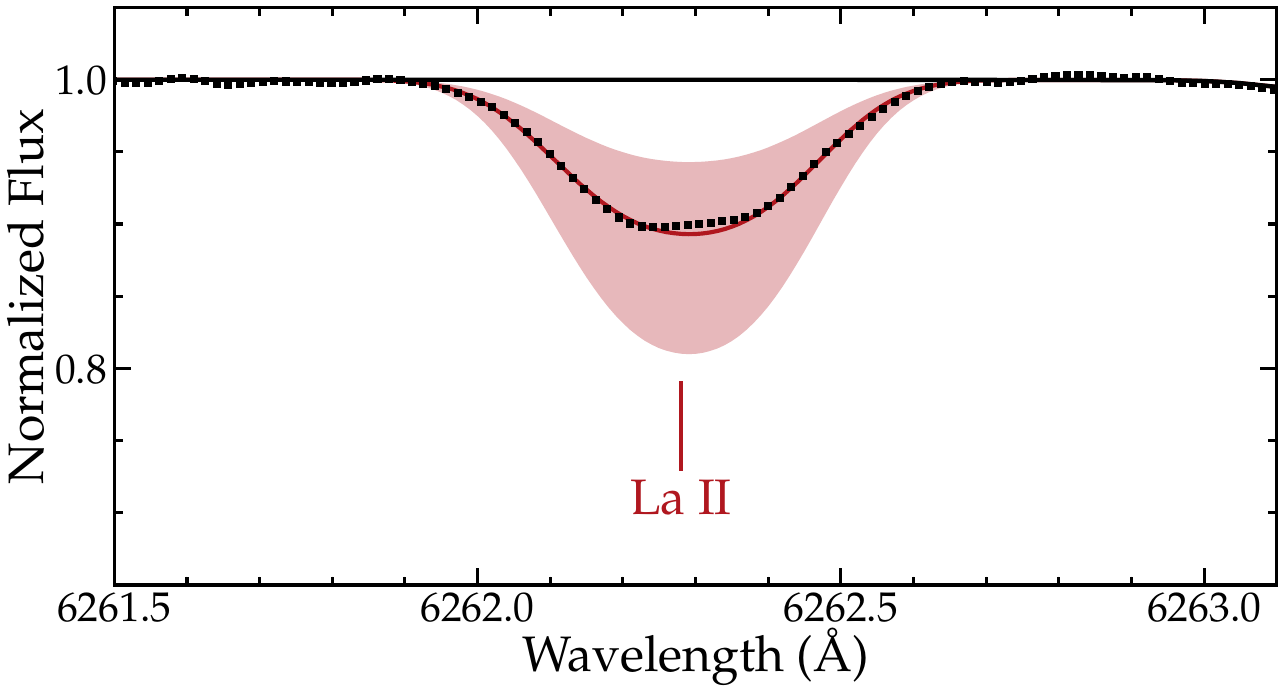}
\hspace*{0.1in}
\includegraphics[angle=0,width=2.5in]{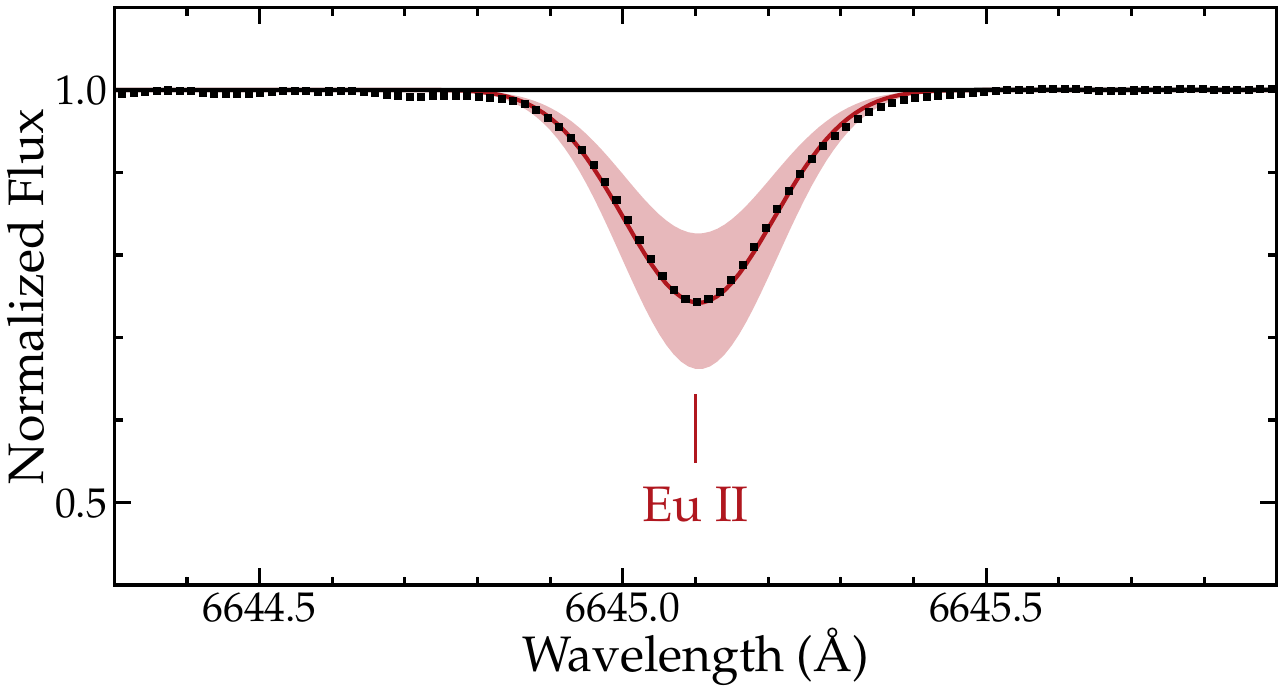} \\
\end{center}
\caption{
\label{fig:synth3}
Comparison of observed spectra (black dots)
and synthetic spectra for several lines of interest.
Symbols are the same as in Figure~\ref{fig:synth1}.
 }
\end{figure*}

Figures~\ref{fig:synth1}--\ref{fig:synth3} 
show fits to several lines of interest.
There are large changes to the 
synthetic spectra induced by changes of $\pm$~0.3~dex
in the input abundance,
demonstrating that
most of these lines are 
on the linear part of the curve-of-growth.
The profile exhibited by lines of
La~\textsc{ii}, 
Ho~\textsc{ii}, and
Lu~\textsc{ii} 
is non-Gaussian.
The seven hyperfine components of the 
Lu~\textsc{ii} $\lambda$5983 line, for example, 
span $\approx$~0.68~\AA.~
Our synthetic line component pattern is based on the
data presented in \citet{lawler09}, and it
provides a satisfactory fit to the observed spectrum.
It is rare for Th~\textsc{ii} lines, other than the one at 4019~\AA,
to be detectable.
Eight Th~\textsc{ii} lines are reliable abundance indicators in \jtwo, 
and four are shown in Figures~\ref{fig:synth1}--\ref{fig:synth3}.
Finally, we detect the U~\textsc{ii} line at 3859~\AA,
as shown in Figure~\ref{fig:synth2},
marking just the eighth time that uranium
has been detected in an \rpro-enhanced star.

\subsection{Kinematics}
\label{sec:kinematics}

We compute the kinematic and orbital properties
of \jtwo.
We adopt the astrometric data from Gaia DR3
and the radial velocity from our MIKE spectra.
We assume that the position and velocity of the Sun are
$(x, y, z)_{\odot}$ = ($-8.178$, $0$, $0$)~kpc 
\citep{gravity19} 
and 
$(v_{x}, v_{y}, v_{z})_{\odot}$ = (11.10, 247.30, 7.25)~\kmsec\
\citep{reid04}.
We draw $10^{5}$ samples of the 
phase-space information 
$(\varpi, \alpha, \delta, v_{r}, \mualpha, \mudelta)$ 
from their associated error distributions,
neglecting the tiny observational uncertainty in 
$(\alpha, \delta)$.
We apply a correction for the zero-point offset in parallax\footnote{%
We use a package 
provided by Gaia Data Processing and Analysis Consortium (DPAC),
available at \url{https://gitlab.com/icc-ub/public/gaiadr3_zeropoint},
to obtain the parallax zero-point offset, $-0.03251$~mas.
}, 
and we account for the correlated observational uncertainties 
in $(\varpi, \mualpha, \mudelta)$. 

We convert these quantities into 
position and velocity vectors in
Galactocentric Cartesian coordinates, 
$\vector{x} = (x,y,z)$ and $\vector{v} = (v_x, v_y, v_z)$, 
in the same manner as \citet{hattori23}.
We derive the kinematic properties of \jtwo\
using the \agama\ package \citep{vasiliev19}.
We calculate the orbital action
$\vector{J} = (J_{r}, J_{z}, J_{\phi})$ 
and energy 
$E=\frac{1}{2}\vector{v}^2 + \Phi_\mathrm{MW}(\vector{x})$ 
using $(\vector{x}, \vector{v})$ 
and adopting the gravitational potential model of the Milky Way,
$\Phi_{\mathrm{MW}}$,
from \cite{mcmillan17}.
We define $J_{\phi} = R v_{\phi} = - (x v_{y} - y v_{x})$, 
so that stars with prograde orbits have $J_{\phi} > 0$ and $v_{\phi} > 0$.
For each realization of $(\vector{x}, \vector{v})$, 
we integrate the orbit forward
in time steps of 1~Myr
for 100~Gyr, which is more than sufficient to 
characterize the orbital properties.
We use these integrations to derive the 
orbital pericentric radius ($r_{\mathrm{peri}}$), 
orbital apocentric radius ($r_{\mathrm{apo}}$), 
maximum excursion above or below the Galactic plane ($z_{\mathrm{max}}$),
and
orbital eccentricity ($e$), defined as
$(r_\mathrm{apo} - r_\mathrm{peri})/(r_\mathrm{apo} + r_\mathrm{peri})$.
Table~\ref{tab:data} summarizes the orbital properties of this star.

\section{Results}
\label{sec:results}

The abundance ratios among the light elements
($Z \leq 30$) in \jtwo\ are typical
for stars with [Fe/H] $= -2.2$.
We briefly summarize these results,
and we refer readers to numerous 
other studies that interpret them
in terms of stellar evolution,
stellar nucleosynthesis, and
Galactic chemical evolution.
Lithium (Li, $Z = 3$) is undetected.
Carbon (C, $Z = 6$) and 
nitrogen (N, $Z = 7$) are not present at
levels that would be considered enhanced.
The $\alpha$ elements O, Mg, Si, Ca, and Ti
($Z = 8$, 12, 14, 20, and 22, respectively)
are enhanced relative to Fe (Fe, $Z = 26$),
[$\alpha$/Fe] $= +0.38 \pm 0.07$.
The ratios of the odd-$Z$ elements sodium, aluminum, and potassium
(Na, Al, and K; $Z = 11$, 13, and 19, respectively)
to Fe are Solar to within a factor of $\approx$~3.
The iron-group elements 
scandium, vanadium, chromium, cobalt, nickel, and zinc
(Sc, V, Cr, Co, Ni, and Zn; 
$Z = 21$, 23, 24, 27, 28, and 30, respectively)
are also present in scaled-Solar proportions.
Only the elements
manganese (Mn, $Z = 25$) and
copper (Cu, $Z = 29$) are present
in sub-Solar abundance ratios, 
[Mn/Fe] $= -0.37 \pm 0.11$ and
[Cu/Fe] $= -0.72 \pm 0.12$.

\begin{figure*}
\begin{center}
\includegraphics[angle=0,width=5.0in]{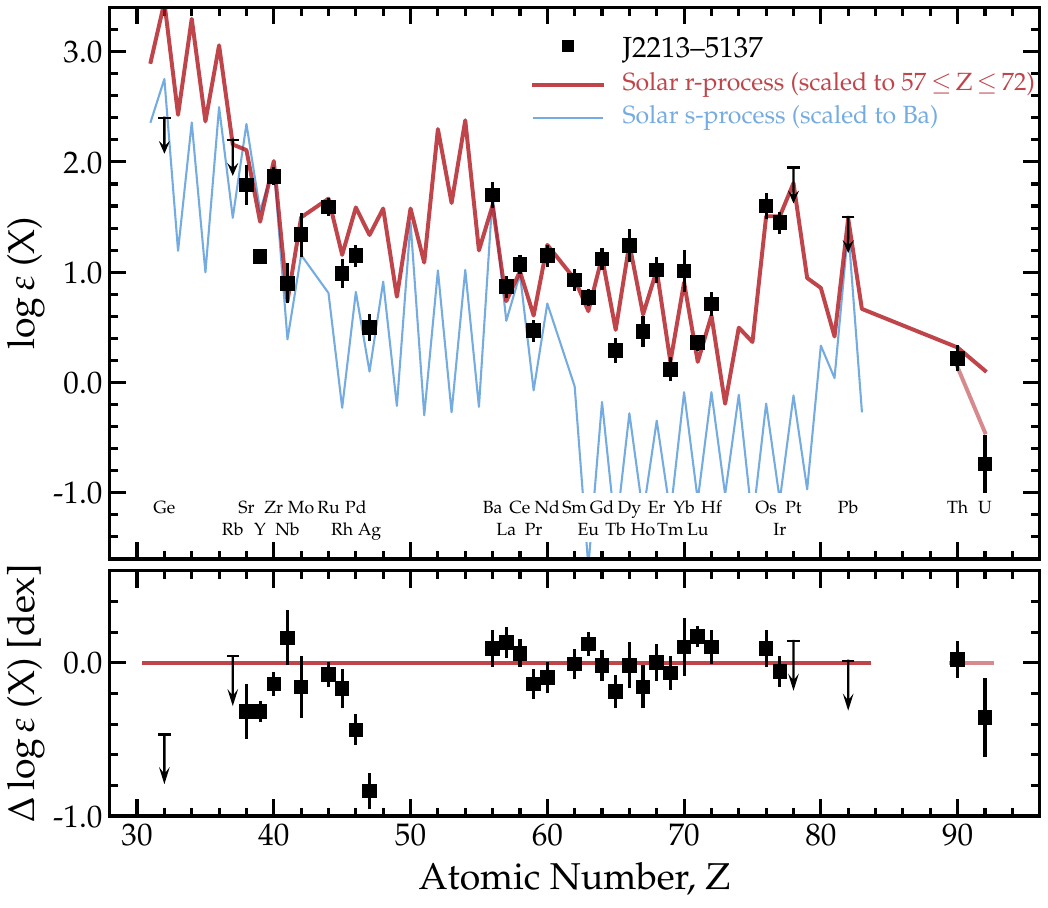}
\end{center}
\caption{
\label{fig:rpropattern}
The heavy-element abundance pattern in \jtwo.
Detections are shown as black squares, and 
upper limits are shown as downward-facing arrows.
The top panel compares the stellar pattern 
with the calculated Solar System \spro\ pattern
and the inferred Solar System \rpro\ residual pattern
(\citealt{sneden08}, except for Y, which is adopted from \citealt{bisterzo14},
and Th and U, which are adopted from \citealt{lodders03}).
The \spro\ pattern is normalized to the Ba abundance, and
the \rpro\ pattern is normalized to the average abundances 
of elements from Ba through Hf.
The light red line connecting Th and U 
accounts for 13~Gyr of radioactive decay, 
which includes an additional 8.5~Gyr relative to the
Solar abundances.
The bottom panel shows the differences relative to the
scaled \rpro\ pattern.
}
\end{figure*}

We detect 29 heavy elements
and
derive upper limits on the abundances of four other heavy elements
not detected in our spectrum of \jtwo.
Figure~\ref{fig:rpropattern} shows
the heavy-element abundance pattern of \jtwo.
The calculated slow neutron-capture process
(\spro) contribution to the Solar System
and the inferred Solar \rpro\ residuals
are shown for comparison.
The Solar \rpro\ pattern provides an excellent fit to the
heavy-element abundance pattern of \jtwo,
whereas the \spro\ one does not.
The Solar \rpro\ pattern and the \jtwo\ pattern
match to within $\approx 2\sigma$, and often much better,
for the elements with $56 \leq Z \leq 82$.
The deviations for Th and U ($Z = 90$ and 92)
are due to their longer radioactive decay ages
relative to the age of the actinides in the Solar System
(see Sections~\ref{sec:uranium} and \ref{sec:ages}).
The lighter \rpro\ elements with $38 \leq Z \leq 47$
exhibit a poor match to both patterns,
and we discuss these elements separately in 
Sections~\ref{sec:rstars} and \ref{sec:fission}.
\jtwo\
exhibits [Eu/Fe] = $+2.45 \pm 0.08$ and
[Ba/Eu] = $-0.73 \pm 0.14$,
identifying it as a member of the class of 
\rthree\ stars proposed by \citet{cain20}.

\section{Discussion}
\label{sec:discussion}

\subsection{No Contamination from Other Processes}
\label{sec:rstars}

Other nucleosynthesis processes could, in principle,
contribute to the heavy-element abundance patterns
in stars with [Fe/H] $> -2.6$ or so \citep{simmerer04}.
In this section, we present evidence that 
any such contamination to \jtwo\ is minimal.

\begin{figure}
\begin{center}
\includegraphics[angle=0,width=3.35in]{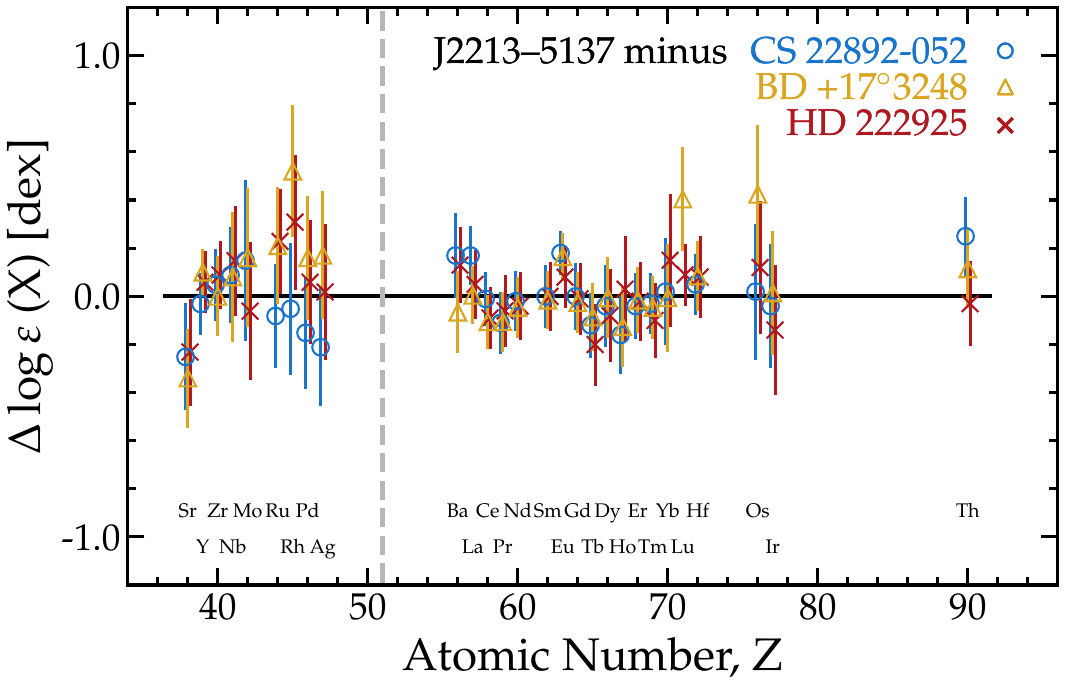}
\end{center}
\caption{
\label{fig:rstars}
Comparison of the heavy-element abundance pattern in \jtwo\
with three other well-studied \rtwo\ stars.
The differences 
for the elements with $Z \geq 56$
are normalized to the lanthanide elements
and their immediate neighbors ($56 \leq Z \leq 72$),
and the differences 
for the elements with $Z \leq 47$
are normalized to the elements with 
$38 \leq Z \leq 42$.
The vertical dashed gray line divides these two mass ranges, and
the horizontal black line indicates a difference of zero.
The points are offset slightly in the horizontal direction
to improve clarity.
References are given in Section~\ref{sec:rstars}.
}
\end{figure}

Three well-studied \rtwo\ stars are 
shown in Figure~\ref{fig:rstars} for comparison.
They have been
selected to span a range of metallicities
and [Eu/Fe] ratios within the \rtwo\ class.
\object[BPS CS 22892-052]{CS~22892-052} was the first
such star identified \citep{sneden94}.
It exhibits [Fe/H] = $-3.12$ and [Eu/Fe] = $+1.65$
\citep{sneden03a,sneden09}.
\object[BD+17 3248]{BD~$+$17$^{\circ}$3248}
exhibits a metallicity similar to \jtwo,
[Fe/H] = $-2.10$
\citep{cowan02}.
Its [Eu/Fe] ratio is near the low end for \rtwo\ stars,
at $+0.90$.
\hdtwo\ is also shown in Figure~\ref{fig:rstars}.
Its metallicity, [Fe/H] = $-1.46$, 
is considerably higher than that of \jtwo,
while its [Eu/Fe] ratio is $+1.32$
\citep{roederer22a}.

Figure~\ref{fig:rstars} demonstrates that there
are no discernible differences among the 
heavy \rpro\ elements and some of the lighter \rpro\ elements
in these four stars.
Differences among the lighter \rpro\ elements
may only be apparent for the elements Ru, Rh, Pd, and Ag
($44 \leq Z \leq 47$).
\citet{roederer23b} identified these four elements as 
being produced, in part, as fission fragments 
of transuranic elements.
Otherwise, 
the similarity among these abundance patterns
across a range of metallicities and levels of \rpro\ enhancement
reinforces two key ideas.
First, these elemental mass ranges appear to exhibit
abundance universality \citep{sneden09,roederer22b},
provided the light and heavy elements are normalized separately.
Secondly, contamination from
other nucleosynthesis processes,
such as charged-particle reactions
or \spro\ nucleosynthesis, is minimal.
Even in relatively metal-rich stars, such as
\jtwo, 
\object[BD+17 3248]{BD~$+$17$^{\circ}$3248}, or even \hdtwo,
the \rpro\ dominates the abundance patterns.

\subsection{Detection of Thorium and Uranium}
\label{sec:uranium}

We detect the actinide elements
thorium (Th, $Z = 90$) and
uranium (U, $Z = 92$) in \jtwo.
Th~\textsc{ii} lines have been detected previously in several dozen
\rpro-enhanced stars
(e.g., \citealt{johnson01th,honda04b,ren12}).
The strongest U~\textsc{ii} lines are weaker
than the strongest Th~\textsc{ii} lines,
and U has only been detected in seven \rpro-enhanced stars previously
\citep{cayrel01,hill02,hill17,cowan02,frebel07he,placco17rpro,holmbeck18,yong21nature}.

\begin{deluxetable*}{ccccccccccccccccc}
\tablecaption{Eu, Pb, Th, and U Abundance Ratios:\ Uncertainties, Producion Ratios, and Inferred Ages
\label{tab:uranium}}
\tablewidth{0pt}
\tabletypesize{\small} 
\tablehead{
\colhead{(1)} &
\colhead{(2)} &
\colhead{(3)} &
\colhead{(4)} &
\colhead{(5)} &
\colhead{(6)} &
\colhead{(7)} &
\colhead{(8)} &
\colhead{(9)} &
\colhead{(10)} &
\colhead{} &
\colhead{(11)} &
\colhead{(12)} &
\colhead{(13)} &
\colhead{(14)} &
\colhead{(15)} &
\colhead{(16)} \\
\hline
\colhead{\logeps{X}} &
\colhead{Value} &
\colhead{$\sigma_{\log gf}$} &
\colhead{$\sigma_{\rm atm}$} &
\colhead{$\sigma_{\rm fit}$} &
\colhead{$\sigma_{\rm total}$} &
\multicolumn{4}{c}{Production Ratios} &
\colhead{} &
\multicolumn{4}{c}{Age} &
\colhead{$\sigma_{\rm age,\,PR}$} &
\colhead{$\sigma_{\rm age,\,obs}$} \\
\cline{7-10} \cline{12-15} \\
\colhead{} &
\colhead{} &
\colhead{} &
\colhead{} &
\colhead{} &
\colhead{} &
\colhead{C02} &
\colhead{S02} &
\colhead{K07} &
\colhead{H17} &
\colhead{} &
\colhead{C02} &
\colhead{S02} &
\colhead{K07} &
\colhead{H17} & 
\colhead{} &
\colhead{}
}
\startdata
Th/Eu  & $-$0.55 &  0.01 &  0.04 &  0.06 &  0.08 & $-$0.29 & $-$0.33 & $-$0.34 & $-$0.24 && 12.0  & 10.3  &   9.9 &  14.6 & 4.5   & 3.7   \\
U/Eu   & $-$1.51 &  0.05 &  0.04 &  0.15 &  0.16 & $-$0.55 & $-$0.55 & $-$0.54 & $-$0.52 && 14.2  & 14.1  &  14.4 &  14.7 & 1.5   & 2.4   \\
U/Th   & $-$0.96 &  0.05 &  0.01 &  0.15 &  0.16 & $-$0.26 & $-$0.22 & $-$0.20 & $-$0.28 && 15.1  & 16.2  &  16.5 &  14.8 & 2.2   & 3.5   \\
Pb/Eu  & $<$0.73 &\nodata&\nodata&\nodata&\nodata& \nodata & \nodata &    0.61 & \nodata &&\nodata&\nodata& (any) &\nodata&\nodata&\nodata\\
Pb/Th  & $<$1.28 &\nodata&\nodata&\nodata&\nodata& \nodata & \nodata &    0.95 & \nodata &&\nodata&\nodata&$<$11.6&\nodata& 4.3   &\nodata\\
Pb/U   & $<$2.24 &\nodata&\nodata&\nodata&\nodata& \nodata & \nodata &    0.81 & \nodata &&\nodata&\nodata&$<$14.2&\nodata& 1.5   &\nodata\\
\hline                                                                                                                                    
Mean   &         &       &       &       &       &         &         &         &         && 13.8  & 13.5  &  13.6 &  14.7 &       &       
\enddata
\tablecomments{%
All ages are reported in units of Gyr.
$\sigma_{\rm age,\,PR}$ is the component of the age uncertainty 
due to an uncertainty in the $\log$ of the production ratio
of 0.1~dex ($\approx 20$~\%).
$\sigma_{\rm age,\,obs}$ is the component of the age uncertainty
due to observational uncertainties (column 6).
}
\tablereferences{%
C02 = \citet{cowan02}, based on \citet{cowan99};
S02 = \citet{schatz02};
K07 = \citet{kratz07}, ``fit2'', including Pb predictions from \citet{roederer09b};
H17 = \citet{hill17}.
}
\end{deluxetable*}

We derive the Th abundance of \jtwo\ from eight Th~\textsc{ii} lines.
Several of these lines are shown in
Figures~\ref{fig:synth1}--\ref{fig:synth3}.
These lines all yield concordant abundances.

We detect the strongest optical U~\textsc{ii} line
at 3859.57~\AA, as shown in Figure~\ref{fig:synth2}.
\citet{shah23} reanalyzed this line
in four stars with previous U detections
and found that it yields concordant abundances with
two additional U~\textsc{ii} lines, 
at 4050.04~\AA\ and 4090.13~\AA.~
In \jtwo,
U~\textsc{ii} accounts for less than $\approx$~5\% of the absorption
at the $\lambda$4050 line, 
which is highly blended with a La~\textsc{ii} line.
The U~\textsc{ii} $\lambda$4090 line is blended with an Fe~\textsc{i} line,
and the total absorption depth is less than 2\% of the continuum.
Continuum placement is uncertain by at least 1\% in this region.
Potential U~\textsc{ii} absorption at 4241.66~\AA\ is also 
coincident with a Zr~\textsc{i} line of uncertain strength.
Upper limits derived from these U~\textsc{ii} lines
and four others that are too weak to detect
(Table~\ref{tab:lines})
are consistent with the U abundance derived from
the $\lambda$3859 line.

Table~\ref{tab:uranium} 
(column~2)
lists the 
\logeps{Th/Eu}, 
\logeps{U/Eu}, 
and 
\logeps{U/Th}
abundance ratios derived from our spectrum of \jtwo.
Eu is traditionally preferred as a stable reference element,
because its production is also dominated by the \rpro\ and
it is readily observable in metal-poor stars.
Table~\ref{tab:uranium} (columns~3--5) lists the 
individual components of the observational uncertainties.
The dominant source of uncertainty is the fitting uncertainty, 
$\sigma_{\rm fit}$, which accounts for
the S/N, continuum placement, and blending features.
Uncertainties in the model atmosphere,
$\sigma_{\rm atm}$,
translate to only small abundance uncertainties when ratios among these
elements are considered, because the abundances are derived
from lines that respond similarly to changes in the 
model atmosphere parameters.
Uncertainties arising from the \loggf\ values,
$\sigma_{\log gf}$,
are virtually negligible,
thanks to extensive laboratory efforts in the last few decades
\citep{lawler01eu,nilsson02u,nilsson02th,ivans06}.
The total uncertainties (column~6)
account for the abundance covariances,
which we calculate using the Monte Carlo resampling method 
described in Section~\ref{sec:abundanalysis},
and these covariances are the reason that the total uncertainties
are smaller than those listed in Table~\ref{tab:abund}.

\jtwo\ exhibits no peculiar behavior among its actinide-element abundances.
It does not exhibit a so-called
``actinide boost'' \citep{hill02,schatz02}.
Its \logeps{Th/Eu} ratio is low,
$-0.55 \pm 0.08$,
and within the range of most other \rpro-enhanced metal-poor stars 
(e.g., \citealt{sneden08,holmbeck18}),
including the \rthree\ star studied by \citet{cain20}.
In contrast,
\object[BPS CS 31082-001]{CS~31082-001}, the original
actinide-boost star, exhibits
\logeps{Th/Eu} $= -0.26 \pm 0.14$ \citep{siqueiramello13}.
We conclude that the Th and U abundances in \jtwo\
are typical among \rpro-enhanced stars.

\subsection{Nuclear Chronometers}
\label{sec:ages}

The age of the \rpro\ material can be calculated from the
radioactive decay of the long-lived 
$^{232}$Th ($t_{1/2}$ = 14.0~Gyr),
$^{235}$U ($t_{1/2}$ = 0.704~Gyr), and
$^{238}$U ($t_{1/2}$ = 4.468~Gyr) isotopes
\citep{fowler60}.
These isotopes can only be produced by \rpro\ nucleosynthesis.
The radioactive decay of
$^{232}$Th, 
$^{235}$U, and
$^{238}$U feeds decay chains that terminate at
$^{208}$Pb,
$^{207}$Pb, and
$^{206}$Pb, respectively.
As the Th and U abundances decrease, the Pb abundance increases.
After 13~Gyr, the Pb abundance will increase by
$\approx$~0.10~dex \citep{roederer09b}.
If the \spro\ contribution to Pb is negligible,
then ratios among
Pb/Eu, Pb/Th, and Pb/U can also be used
to calculate the age 
\citep{clayton64}.

This method can be applied when production of the \rpro\ material was
dominated by a single event (e.g., \citealt{cowan97,cowan99,cayrel01}).
Otherwise, a chemical evolution model is required
(e.g., \citealt{thielemann83,cowan87,mathews88}).
The high level of \rpro\ enhancement leaves little doubt that
the single-event scenario can be applied to \jtwo.
A modest amount of caution is appropriate, however, 
because previous studies of a 
few actinide boost stars have found that the levels of the actinide elements
can be so enhanced that this method returns negative, non-physical ages
(e.g., \citealt{hill02, mashonkina14he}),
and there are hints of a potential metallicity dependence
\citep{saraf23}.

This method also requires that the initial production ratios
are known.
Table~\ref{tab:uranium} (columns~7--10)
lists the production ratios
predicted by several methods.
These calculations are frequently tuned to reproduce the 
Solar System \rpro\ isotopic component, so they are not
independent, but their range provides an estimate
of their level of uncertainty.
The predicted initial production ratios 
are also sensitive to the choice of nuclear physics
inputs and the \rpro\ model (e.g., \citealt{cowan99,holmbeck19a}).
\citet{schatz02} estimated that the uncertainties in the
production ratios may be $\approx$~0.10--0.12~dex.
Table~\ref{tab:uranium} (column~15) lists the
component of the age uncertainty that results from
an uncertainty of 0.10~dex in the production ratios.

The absolute ages, calculated using equations~1--3 of
\citet{cayrel01}
and Equations~\ref{eqn:pb1}--\ref{eqn:pb5}
in Appendix~\ref{sec:appendix},
are listed in Table~\ref{tab:uranium}
(columns~11-14).
Table~\ref{tab:uranium} (column~16) lists the 
component of the age uncertainty that results from
the observational uncertainty in each abundance ratio
(Section~\ref{sec:uranium}).
We resample the observed abundances and the production ratios
and compute the mean of the resulting age distributions.
These values, which range from 13.5~Gyr to 14.7~Gyr
for different sets of production ratios,
are listed in the bottom row of Table~\ref{tab:uranium}.
Their uncertainties, computed from the 16th and 84th percentiles
of the distributions, are $\approx$~2.6~Gyr.

The Pb abundance in \jtwo\ is an upper limit only,
so the ages predicted from ratios involving the 
radiogenic production of Pb through actinide decay
are also upper limits.
These limits are consistent with the
ages derived from the Th/Eu, U/Eu, and U/Th ratios.
Detecting the Pb~\textsc{ii} line at 2203~\AA,
which should be strong in \jtwo,
would provide a unique opportunity to calculate ages from 
both the parent and daughter nuclei
in a star where all three abundances are derived from
lines of the majority species \citep{roederer20}.

\begin{figure}
\begin{center}
\includegraphics[angle=0,width=3.35in]{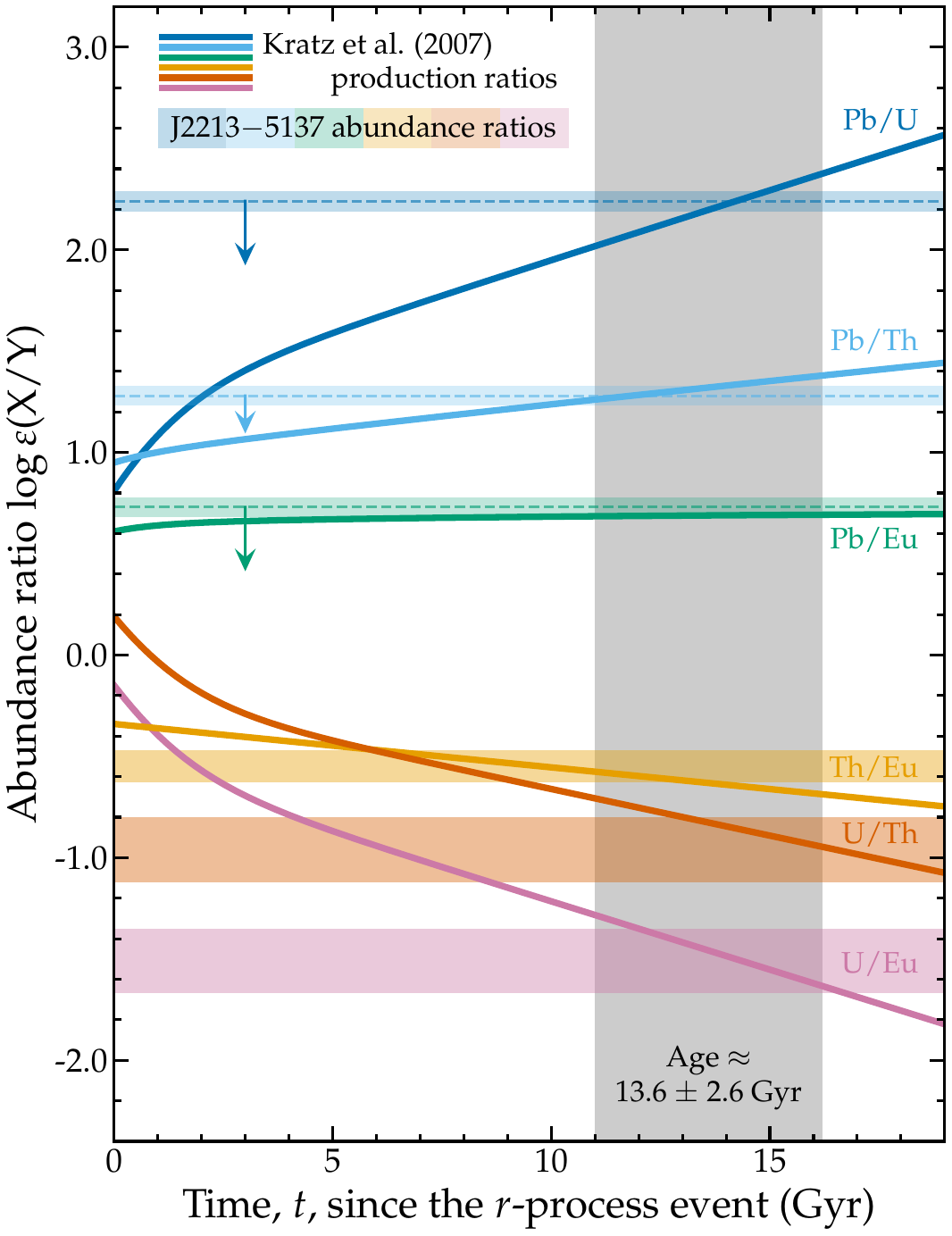}
\end{center}
\caption{
\label{fig:ages}
Comparison of observed (shaded horizontal bands)
and predicted (solid lines) abundance ratios
as a function of age for \jtwo.
The production ratios are drawn from \citet{kratz07},
supplemented with Pb data presented in \citet{roederer09b}
from the same set of calculations.
The \logeps{U/Eu} and \logeps{U/Th} curves 
include the $+$0.39~dex contribution from
the relatively rapid decay of 
$^{235}$U, which is otherwise not accounted for
in equations~2 and 3 of \citet{cayrel01}.
Those equations only accounted for $^{238}$U decay, and
the \logeps{$^{238}$U/Eu} and
\logeps{$^{238}$U/Th} contributions have production
ratios of $-0.54$ and $-0.20$, respectively, at time $t = 0$, 
which are the values listed in Table~\ref{tab:uranium}.
The equations relating the Pb abundance to age
are presented in Appendix~\ref{sec:appendix}.
The age (vertical gray band) is computed based on the
Th/Eu, U/Th, and U/Eu ratios.
}
\end{figure}

Figure~\ref{fig:ages} compares the predicted abundance ratios
involving Eu, Pb, Th, and U with the 
ratios derived for \jtwo.
To simplify the figure, only one set of predicted production ratios
is shown.
The age inferred from the \citet{kratz07}
production ratios, $13.6 \pm 2.6$~Gyr, is shown
by the vertical band in Figure~\ref{fig:ages}.

\subsection{Transuranic Fission Fragments}
\label{sec:fission}

\begin{figure*}
\begin{center}
\includegraphics[angle=0,width=5.5in]{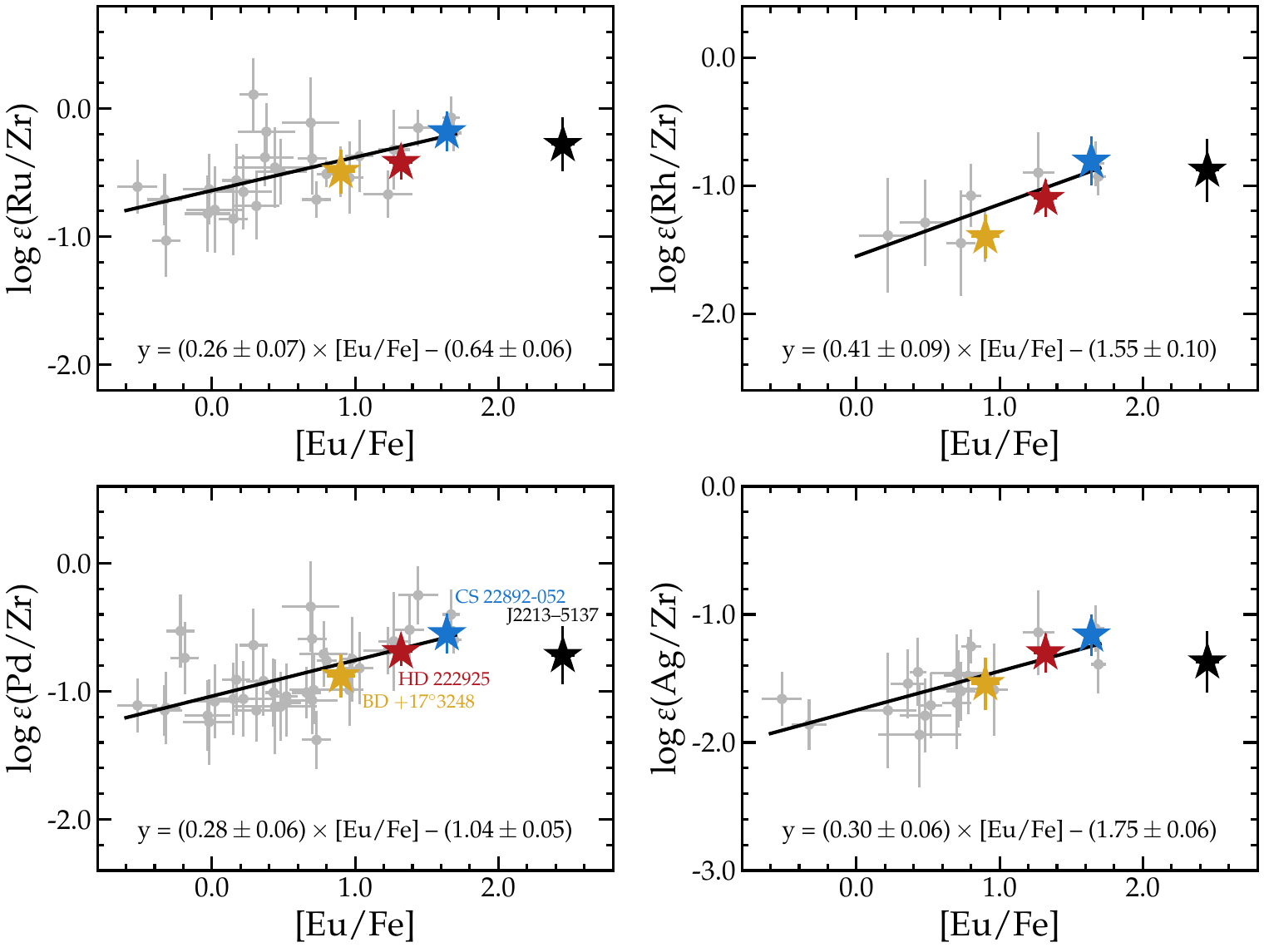}
\end{center}
\caption{
\label{fig:dilute_ratios}
Ratios most affected by fission-fragment distributions
for \jtwo\ and the sample of stars studied by \citet{roederer23b}.
A few stars are marked in separate colors and labeled individually.
The black line marks a best fit to the 
sample of stars studied by \citeauthor{roederer23b},
excluding \jtwo.
The equation for this fiducial line is provided in each panel,
where \textit{y} is shorthand for the 
$\log\varepsilon$(X/Zr) ratio printed on the vertical axis in each panel.
}
\end{figure*}

\citet{roederer23b} noticed that the
\logeps{Ru/Zr}, 
\logeps{Rh/Zr},
\logeps{Pd/Zr}, and
\logeps{Ag/Zr} ratios in \rpro-enhanced stars
correlated with the [Eu/Fe] ratio.
That study interpreted those correlations as potential evidence
that fission fragments of transuranic elements
boosted the production of Ru, Rh, Pd, and Ag in 
strong \rpro\ events that might produce more Eu, 
lanthanides, and transuranic elements.
Figure~\ref{fig:dilute_ratios} shows the
relationships between these elements for the sample of stars
examined by \citeauthor{roederer23b} 
\jtwo\ is also shown in Figure~\ref{fig:dilute_ratios}, and
it extends the range of [Eu/Fe] ratios to much higher values
than were available previously.
\jtwo\ does not continue the upward extrapolation
of the trends identified by \citeauthor{roederer23b} 
Instead, \jtwo\ exhibits relative enhancements of Ru, Rh, Pd, and Ag
that are similar to those found in stars
with $+1.3 \leq$ [Eu/Fe] $\leq +1.7$ or so.

Perhaps the fact that these trends do not continue to higher [Eu/Fe] ratios
suggests that there is a maximum strength of the \rpro.
For example, there could be 
a limit on the number of times fission cycling occurs, 
or there could be a maximum duration of time when the neutron density 
remains high enough for an \rpro.
Studying the four key elements---Ru, Rh, Pd, and Ag---%
in additional \rpro-enhanced stars with [Eu/Fe] $> +1.7$
will be critical to understand this behavior.
Among these elements, only Ru was
derived by \citet{cain20} for 
\cainstar, the \rthree\ star known previously.
The high value of its \logeps{Ru/Zr} ratio,
$-0.24 \pm 0.24$,
is within the same range as found for
\jtwo\ and other stars with [Eu/Fe] $> +1.3$.
Abundances of these elements have not been published
for any other \rpro-enhanced star with
[Eu/Fe] $> +1.7$.
We predict that \rpro-enhanced stars in the \rettwolong\ dwarf galaxy,
which exhibit 
[Eu/Fe] = $+1.69 \pm 0.12$ or
[Eu/H] = $-0.98 \pm 0.13$ on average \citep{ji16ret2},
should exhibit similarly high
\logeps{Ru,Rh,Pd,Ag/Zr} ratios.

\subsection{A New Method to Estimate the Dilution Mass or \rpro\ Yields}
\label{sec:dilution}

We propose that the horizontal dispersion of stars around the best-fit
fiducial lines in Figure~\ref{fig:dilute_ratios} may reflect a combination
of dilution into various amounts of Fe in the
interstellar medium (ISM) and
a range of \rpro\ yields from the events that
enriched each of the stars in the sample.
We assume that the abundance ratio shown on the vertical axis in each panel
is a fixed, intrinsic property of an \rpro\ nucleosynthesis event.
If the fiducial line in each panel
reveals the average behavior of the sample,
then each star's horizontal displacement in [Eu/Fe]
reflects the \textit{relative} variation in dilution
(i.e., the Fe abundance varies) 
and/or yields 
(i.e., the Eu abundance varies).

We express the relative displacement in [Eu/Fe], \deltaeufe, as follows:\
\begin{eqnarray}
\label{eqn:dilution}
\log \delta_{\rm EuFe} &=& 
            {\rm [Eu/Fe]}_{\rm excess} \nonumber \\
        &=& {\rm [Eu/Fe]}_{\star} - {\rm [Eu/Fe]_{\rm fiducial}} \nonumber \\
        &=& {\rm [Eu/Fe]}_{\star} - (\log\varepsilon({\rm X}/{\rm Zr}) - b)/a.
\end{eqnarray}
Here, 
X is one of the elements that exhibits this behavior:\ Ru, Rh, Pd, or Ag.
The constants $a$ and $b$ are the slope and intercept, respectively, of the
best-fit line to the relationships shown in Figure~\ref{fig:dilute_ratios},
which were derived in \citet{roederer23b}.
The dilution and yields are degenerate, so 
multiple combinations of dilution values and yield variations
are permissible.
We emphasize that \deltaeufe\ reflects the 
dilution or \rpro\ yields relative to the rest of the sample;
it is not an absolute measure.
For example, consider a hypothetical star for which [Eu/Fe]
is 0.3~dex higher than the fiducial
[Eu/Fe] ratio that corresponds to the observed \logeps{Pd/Zr} ratio.
Assume that this \rpro\ material 
was diluted into an average amount of Fe in the ISM.~
Equation~\ref{eqn:dilution} would thus predict that
the \rpro\ event that enriched this star had a yield that was
$10^{\delta_{\rm EuFe}} = 10^{0.3} = 2$
times higher than an \rpro\ event
that enriched an average star in the sample.

\begin{figure}
\begin{center}
\includegraphics[angle=0,width=3.35in]{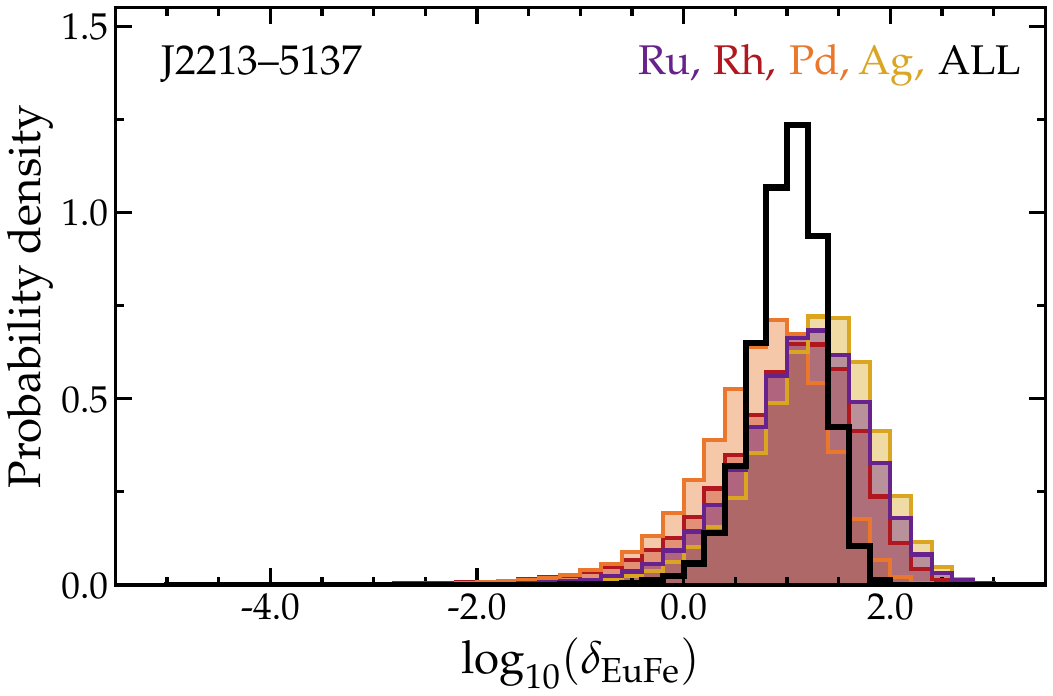} \\
\vspace*{0.05in}
\includegraphics[angle=0,width=3.35in]{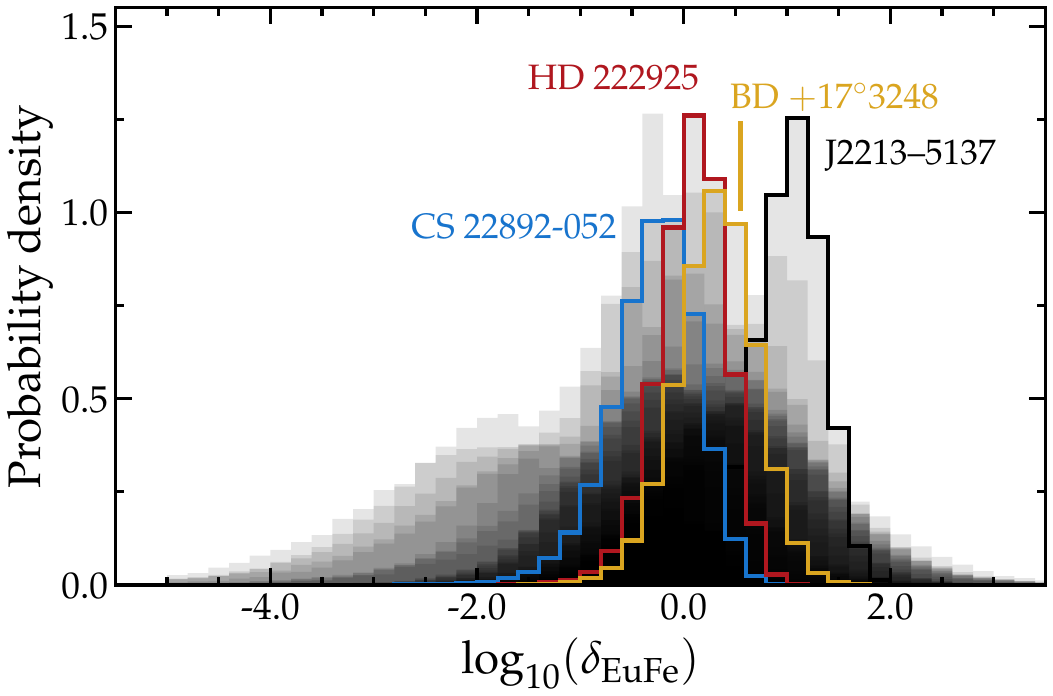} \\
\end{center}
\caption{
\label{fig:dilution}
Probability distribution densities for \deltaeufe\
predicted using Equation~\ref{eqn:dilution}
for the individual element ratios \logeps{X/Zr}
in \jtwo\ (top panel)
and probabilities combined from up to 
four individual element ratios
found in 41~stars in the \citet{roederer23b} sample (bottom panel).
There are 42~stars in the \citeauthor{roederer23b}\ sample,
but no abundances of Ru, Rh, Pd, or Ag have been reported for
one of them,
\object[HD 140283]{HD~140283}.
The distribution for each star is shown by a light gray region.
By construction, they are
centered around $\log$\,\deltaeufe = 0.0.
A few stars of note are labeled.
}
\end{figure}

We next estimate \deltaeufe\ for \jtwo.
We draw $10^{5}$ resamples from the 
[Eu/Fe], 
\logeps{Zr},
\logeps{Ru}, 
\logeps{Rh},
\logeps{Pd}, and
\logeps{Ag} 
abundances and their uncertainties,
as well as the fiducial lines and their uncertainties.
For each resample, we compute the \logeps{X/Zr} ratios
and \deltaeufe.
The top panel of Figure~\ref{fig:dilution}
shows the relative \deltaeufe\ parameters
predicted by each of the four abundance ratios.
We resample each of these four distributions 
and compute the mean of those four resamples $10^{5}$ times
to estimate the combined relative \deltaeufe\ parameter.
That value,
$\log$\,\deltaeufe $= +1.04_{-0.35}^{+0.30}$,
is precise to within a factor of $\approx$~2.
We repeat this set of calculations for 
41~stars from \citet{roederer23b}.
The results are
listed in Table~\ref{tab:dilution} and
shown in the bottom panel of Figure~\ref{fig:dilution}.

\begin{deluxetable*}{lccccccc}
\tablecaption{Relative $\delta_{\rm EuFe}$ Parameters
for 42 Stars, Sorted by Decreasing [Eu/Fe] Ratios
\label{tab:dilution}}
\tablewidth{0pt}
\tabletypesize{\small} 
\tablehead{
\colhead{Star} &
\colhead{[Eu/Fe]} &
\colhead{[Fe/H]} &
\colhead{$\log$ \deltaeufe} &
\colhead{$\log$ \deltaeufe} &
\colhead{$\log$ \deltaeufe} &
\colhead{$\log$ \deltaeufe} &
\colhead{$\log$ \deltaeufe} \\
\colhead{} &
\colhead{} &
\colhead{} &
\colhead{\logeps{Ru/Zr}} &
\colhead{\logeps{Rh/Zr}} &
\colhead{\logeps{Pd/Zr}} &
\colhead{\logeps{Ag/Zr}} &
\colhead{combined} 
}
\startdata
2MASS J22132050$-$5137385    & $+$2.45   & $-$2.20 & $+$1.06$^{+0.55}_{-0.75}$ & $+$0.82$^{+0.51}_{-0.68}$ & $+$1.30$^{+0.52}_{-0.62}$ & $+$1.18$^{+0.54}_{-0.65}$ & $+$1.04$^{+0.30}_{-0.35}$ \\
CS 31082-001                 & $+$1.69   & $-$2.90 & $-$0.04$^{+0.67}_{-0.92}$ & $+$0.18$^{+0.49}_{-0.63}$ & $+$0.12$^{+0.48}_{-0.63}$ & $+$0.49$^{+0.78}_{-0.88}$ & $+$0.12$^{+0.35}_{-0.40}$ \\
CS 29497-004                 & $+$1.67   & $-$2.84 & $-$0.52$^{+0.77}_{-1.11}$ & $-$0.08$^{+0.56}_{-0.74}$ & $-$0.62$^{+0.78}_{-1.00}$ & $-$0.47$^{+0.71}_{-0.88}$ & $-$0.50$^{+0.40}_{-0.49}$ \\
CS 22892-052                 & $+$1.64   & $-$3.09 & $-$0.13$^{+0.71}_{-0.97}$ & $-$0.16$^{+0.59}_{-0.77}$ & $-$0.11$^{+0.63}_{-0.81}$ & $-$0.33$^{+0.64}_{-0.80}$ & $-$0.26$^{+0.36}_{-0.44}$ \\
2MASS J14325334$-$4125494    & $+$1.44   & $-$2.97 & $-$0.45$^{+0.69}_{-0.98}$ & \nodata                   & $-$1.38$^{+0.94}_{-1.20}$ & \nodata                   & $-$0.98$^{+0.63}_{-0.80}$ \\
HE 1219$-$0312               & $+$1.38   & $-$2.97 & \nodata                   & \nodata                   & $-$0.48$^{+0.99}_{-1.17}$ & \nodata                   & $-$0.48$^{+0.99}_{-1.17}$ \\
HD 222925                    & $+$1.32   & $-$1.46 & $+$0.47$^{+0.58}_{-0.71}$ & $+$0.22$^{+0.46}_{-0.56}$ & $+$0.07$^{+0.48}_{-0.59}$ & $-$0.18$^{+0.58}_{-0.70}$ & $+$0.10$^{+0.29}_{-0.34}$ \\
2MASS J15383085$-$1804242    & $+$1.27   & $-$2.09 & $+$0.04$^{+1.23}_{-1.42}$ & $-$0.32$^{+0.83}_{-0.97}$ & $-$0.27$^{+1.40}_{-1.53}$ & $-$0.76$^{+1.13}_{-1.30}$ & $-$0.39$^{+0.62}_{-0.69}$ \\
CS 31078-018                 & $+$1.23   & $-$2.84 & $+$1.34$^{+0.81}_{-0.78}$ & \nodata                   & $-$0.06$^{+0.71}_{-0.83}$ & \nodata                   & $+$0.63$^{+0.56}_{-0.58}$ \\
CS 22953-003                 & $+$1.03   & $-$2.84 & $-$0.02$^{+1.12}_{-1.26}$ & \nodata                   & $+$0.25$^{+1.03}_{-1.11}$ & \nodata                   & $+$0.09$^{+0.78}_{-0.86}$ \\
HE 2327$-$5642               & $+$0.98   & $-$2.79 & \nodata                   & \nodata                   & $-$0.06$^{+1.21}_{-1.29}$ & \nodata                   & $-$0.06$^{+1.21}_{-1.29}$ \\
CS 22896-154                 & $+$0.96   & $-$2.69 & $+$0.57$^{+1.12}_{-1.18}$ & \nodata                   & $+$0.78$^{+1.04}_{-1.06}$ & $+$0.42$^{+1.23}_{-1.26}$ & $+$0.58$^{+0.69}_{-0.70}$ \\
BD $+$17$^{\circ}$3248       & $+$0.90   & $-$2.10 & $+$0.32$^{+0.80}_{-0.88}$ & $+$0.54$^{+0.55}_{-0.59}$ & $+$0.33$^{+0.62}_{-0.68}$ & $+$0.20$^{+0.72}_{-0.78}$ & $+$0.33$^{+0.36}_{-0.38}$ \\
HD 221170                    & $+$0.80   & $-$2.20 & $+$0.30$^{+0.46}_{-0.54}$ & $-$0.35$^{+0.66}_{-0.77}$ & $-$0.20$^{+0.43}_{-0.52}$ & $-$0.87$^{+0.54}_{-0.68}$ & $-$0.32$^{+0.29}_{-0.33}$ \\
CD $-$45$^{\circ}$3283       & $+$0.78   & $-$0.99 & \nodata                   & \nodata                   & $-$0.40$^{+0.95}_{-1.04}$ & $-$0.12$^{+1.04}_{-1.11}$ & $-$0.28$^{+0.73}_{-0.77}$ \\
HD 20                        & $+$0.73   & $-$1.60 & $+$1.00$^{+0.64}_{-0.60}$ & $+$0.48$^{+1.04}_{-1.06}$ & $+$1.94$^{+0.96}_{-0.83}$ & $+$0.23$^{+0.79}_{-0.83}$ & $+$0.92$^{+0.45}_{-0.44}$ \\
HD 120559                    & $+$0.71   & $-$1.31 & \nodata                   & \nodata                   & $+$0.53$^{+0.99}_{-1.02}$ & $+$0.14$^{+1.05}_{-1.08}$ & $+$0.33$^{+0.74}_{-0.76}$ \\
HD 3567                      & $+$0.70   & $-$1.33 & \nodata                   & \nodata                   & $+$0.81$^{+0.99}_{-0.98}$ & $-$0.27$^{+1.04}_{-1.12}$ & $+$0.26$^{+0.74}_{-0.75}$ \\
BS 17569-049                 & $+$0.70   & $-$2.88 & $-$0.26$^{+1.11}_{-1.26}$ & \nodata                   & $-$0.90$^{+1.04}_{-1.18}$ & $+$0.51$^{+1.23}_{-1.24}$ & $-$0.25$^{+0.69}_{-0.73}$ \\
2MASS J18301354$-$4555101    & $+$0.69   & $-$3.57 & $-$1.35$^{+1.39}_{-1.72}$ & \nodata                   & $-$1.81$^{+1.31}_{-1.54}$ & \nodata                   & $-$1.65$^{+1.02}_{-1.18}$ \\
HD 115444                    & $+$0.66   & $-$2.85 & \nodata                   & \nodata                   & $+$0.55$^{+0.76}_{-0.77}$ & \nodata                   & $+$0.55$^{+0.76}_{-0.77}$ \\
HD 186478                    & $+$0.52   & $-$2.60 & \nodata                   & \nodata                   & $+$0.52$^{+0.95}_{-0.95}$ & $+$0.39$^{+0.88}_{-0.90}$ & $+$0.45$^{+0.66}_{-0.67}$ \\
HD 6268                      & $+$0.52   & $-$2.63 & \nodata                   & \nodata                   & $+$0.70$^{+0.98}_{-0.97}$ & \nodata                   & $+$0.70$^{+0.98}_{-0.97}$ \\
HD 108317                    & $+$0.48   & $-$2.37 & $-$0.10$^{+1.03}_{-1.11}$ & $-$0.15$^{+0.87}_{-0.94}$ & $+$0.77$^{+1.03}_{-1.01}$ & $+$0.61$^{+1.02}_{-1.01}$ & $+$0.27$^{+0.52}_{-0.54}$ \\
HD 160617                    & $+$0.44   & $-$1.77 & $-$0.26$^{+1.27}_{-1.38}$ & \nodata                   & $+$0.73$^{+1.40}_{-1.39}$ & $+$1.07$^{+1.46}_{-1.42}$ & $+$0.51$^{+0.83}_{-0.84}$ \\
BD $+$08$^{\circ}$2856       & $+$0.43   & $-$2.11 & \nodata                   & \nodata                   & $+$0.32$^{+0.98}_{-0.99}$ & $-$0.56$^{+0.90}_{-0.99}$ & $-$0.14$^{+0.69}_{-0.71}$ \\
HD 84937                     & $+$0.38   & $-$2.25 & $-$1.39$^{+0.93}_{-1.20}$ & \nodata                   & \nodata                   & \nodata                   & $-$1.39$^{+0.93}_{-1.20}$ \\
HD 19445                     & $+$0.37   & $-$2.15 & $-$0.63$^{+0.91}_{-1.06}$ & \nodata                   & \nodata                   & \nodata                   & $-$0.63$^{+0.91}_{-1.05}$ \\
HD 108577                    & $+$0.36   & $-$2.36 & \nodata                   & \nodata                   & $-$0.07$^{+0.99}_{-1.02}$ & $-$0.34$^{+0.92}_{-0.98}$ & $-$0.22$^{+0.69}_{-0.71}$ \\
HD 107752                    & $+$0.31   & $-$2.85 & $+$0.78$^{+1.13}_{-1.07}$ & \nodata                   & $+$0.70$^{+0.93}_{-0.90}$ & \nodata                   & $+$0.75$^{+0.76}_{-0.72}$ \\
CS 22966-057                 & $+$0.29   & $-$2.62 & $-$2.59$^{+1.20}_{-1.63}$ & \nodata                   & $-$1.15$^{+1.03}_{-1.15}$ & \nodata                   & $-$1.93$^{+0.84}_{-1.03}$ \\
HD 126238                    & $+$0.22   & $-$1.93 & $+$0.25$^{+1.22}_{-1.21}$ & $-$0.17$^{+1.14}_{-1.19}$ & $+$0.29$^{+1.11}_{-1.10}$ & $+$0.22$^{+1.54}_{-1.56}$ & $+$0.14$^{+0.67}_{-0.67}$ \\
HD 2796                      & $+$0.17   & $-$2.47 & $-$0.14$^{+1.14}_{-1.17}$ & \nodata                   & $-$0.29$^{+1.03}_{-1.09}$ & \nodata                   & $-$0.23$^{+0.79}_{-0.81}$ \\
CS 29518-051                 & $+$0.15   & $-$2.69 & $+$1.00$^{+1.24}_{-1.11}$ & \nodata                   & $+$0.22$^{+1.06}_{-1.05}$ & \nodata                   & $+$0.62$^{+0.84}_{-0.79}$ \\
HD 85773                     & $+$0.02   & $-$2.62 & $+$0.60$^{+1.43}_{-1.35}$ & \nodata                   & $+$0.16$^{+1.10}_{-1.09}$ & \nodata                   & $+$0.40$^{+0.92}_{-0.90}$ \\
HD 128279                    & $-$0.02   & $-$2.46 & $-$0.06$^{+1.14}_{-1.14}$ & \nodata                   & $+$0.69$^{+1.30}_{-1.22}$ & \nodata                   & $+$0.32$^{+0.89}_{-0.86}$ \\
HD 110184                    & $-$0.03   & $-$2.52 & $+$0.66$^{+1.27}_{-1.18}$ & \nodata                   & $+$0.50$^{+1.07}_{-1.02}$ & \nodata                   & $+$0.60$^{+0.86}_{-0.80}$ \\
CS 22873-055                 & $-$0.19   & $-$2.99 & \nodata                   & \nodata                   & $-$1.25$^{+1.01}_{-1.14}$ & \nodata                   & $-$1.25$^{+1.01}_{-1.14}$ \\
BD $-$18$^{\circ}$5550       & $-$0.22   & $-$3.06 & \nodata                   & \nodata                   & $-$2.04$^{+1.04}_{-1.20}$ & \nodata                   & $-$2.04$^{+1.04}_{-1.20}$ \\
CS 22873-166                 & $-$0.32   & $-$2.97 & $+$1.18$^{+1.34}_{-1.12}$ & \nodata                   & $-$0.00$^{+1.07}_{-1.04}$ & \nodata                   & $+$0.62$^{+0.87}_{-0.80}$ \\
HD 88609                     & $-$0.33   & $-$3.07 & $-$0.06$^{+0.86}_{-0.83}$ & \nodata                   & $+$0.06$^{+0.78}_{-0.75}$ & $+$0.03$^{+0.74}_{-0.70}$ & $+$0.02$^{+0.48}_{-0.46}$ \\
HD 122563                    & $-$0.52   & $-$2.77 & $-$0.64$^{+0.88}_{-0.90}$ & \nodata                   & $-$0.27$^{+0.81}_{-0.80}$ & $-$0.82$^{+0.75}_{-0.77}$ & $-$0.58$^{+0.49}_{-0.50}$ \\
\enddata
\end{deluxetable*}

The peaks of the distributions shown in Figure~\ref{fig:dilution}
span nearly 3~dex in 
$\log$\,\deltaeufe.
In other words, 
the combined effects of dilution and \rpro\ yields
that contributed to this sample of stars
span three orders of magnitude.
The long tail to low \deltaeufe\ values
reflects the large uncertainties in \deltaeufe\ 
that result from few and uncertain measurements of Ru, Rh, Pd, and Ag
in stars with overall low abundances of these elements.
\jtwo\ and a few stars of note (Section~\ref{sec:rstars})
are labeled
in Figure~\ref{fig:dilution}.
\jtwo\ marks the low-dilution or high-yield 
extreme of the sample.
We conclude that
the \rpro\ material from which \jtwo\ formed
was mixed into an ISM that was $\approx$~11 times (i.e., $10^{+1.04}$) 
less abundant in Fe than the typical star in the sample,
or that the \rpro\ event that enriched \jtwo\
ejected $\approx$~11 times more \rpro\ material
than the typical event,
or some combination of these two scenarios.

\subsection{Exploring a Potential Dwarf Galaxy Connection}
\label{sec:dwarfgalaxy}

We investigate three potential connections between
\jtwo\ and stars born in dwarf galaxies.

First,
the age estimate for \jtwo\ calculated in Section~\ref{sec:ages},
$13.6 \pm 2.6$~Gyr,
matches
the epoch of star formation in low-mass dwarf galaxies
\citep{brown14,weisz14,sacchi21}
and the oldest globular clusters
\citep{vandenberg13,ying23}.
A few stars that were likely formed in globular clusters
are known in the halo
(e.g., \citealt{ramirez12gc,lind15,martell16}),
but we do not consider \jtwo\ to be one of them.
\jtwo\ does not exhibit any of the abundance variations
among light elements---such as O, Na, Mg, or Al---%
found in many stars in globular clusters
(e.g., \citealt{carretta09uves}).
It is therefore unlikely to have been formed as a member of the
second, or chemically polluted, generation of stars formed 
within a globular cluster.
Furthermore, the level of \rpro\ enhancement is far in excess
of any known stars within globular clusters 
(e.g., \citealt{gratton04,cabreragarcia24}).
In contrast, as noted in Section~\ref{sec:intro}, 
stars in dwarf galaxies exhibit abundance patterns
similar to that found in \jtwo.
This line of reasoning does not require a dwarf galaxy origin,
but it is compatible with one.

Secondly,
\jtwo\ is bound to the inner region of the Milky Way.
It has an eccentric orbit ($e$ = 0.73)
with a small orbital pericenter ($r_{\mathrm{peri}}$ = 1.1~kpc).
It is currently ($R$ = 7.0~kpc)
near its orbital apocenter ($r_{\mathrm{apo}}$ = 7.2~kpc),
which is within the Sun's orbital radius.
These orbital characteristics
are broadly similar to other \rtwo\ stars in the Solar neighborhood 
\citep{roederer18d,gudin21,hattori23},
but the orbit of \jtwo\ is distinct from them.
For example, 
the orbit of
the other \rthree\ field star, \cainstar,
is much wider:\
$r_{\mathrm{peri}} \approx$~6~kpc and
$r_{\mathrm{apo}} \approx$~26~kpc \citep{cain20,hattori23}.
The orbit, actions, and energy of \jtwo\
differ from all stars
in the \citeauthor{hattori23}\ sample with [Eu/Fe] $> +1.7$.
We conclude that it is unlikely that \jtwo\ formed alongside
any of these stars.

The orbital properties of \jtwo\ overlap with 
the \mbox{H22:DTC-15}
chemodynamically tagged group of \rpro-enhanced stars,
which was identified by \citet{hattori23}.
The 18 members of this group have orbital pericenters
$< 1.2$~kpc (mostly $< 0.6$~kpc), 
orbital apocenters of 6--12~kpc (mostly $> 8$~kpc), 
and radial but slightly prograde orbits.
These stars exhibit a mean metallicity of
$-2.43$ and a metallicity dispersion of 0.50~dex,
and they include two of the comparison stars
considered in Section~\ref{sec:rstars},
\object[BD+17 3248]{BD~$+$17$^{\circ}$3248} and
\object[BPS CS 22892-052]{CS~22892-052}.
Figure~\ref{fig:rstars} compares the 
heavy-element abundance patterns of \jtwo\
and these two stars.
Most of their abundance ratios are indistinguishable.
As noted previously, however,
the \logeps{Ru,Rh,Pd,Ag/Zr} ratios are higher in 
\object[BPS CS 22892-052]{CS~22892-052} than in \jtwo\ by about 0.1~dex,
and 
these same ratios are lower in
\object[BD+17 3248]{BD~$+$17$^{\circ}$3248} by about 0.2~dex.
This difference implies that
the \rpro\ material found in these three stars
could not have originated in only one \rpro\ event.
Two or more \rpro\ events could have enriched the 
putative progenitor dwarf galaxy,
which is unlikely because of the 
low occurrence frequency of \rpro-enhanced dwarf galaxies today
\citep{ji16nat}.
Alternatively,
\mbox{H22:DTC-15} could include stars from several 
accretion events, and therefore
\jtwo\ may not actually share an origin with
either of these stars.
This association should be reassessed
once larger samples of \rpro-enhanced stars have been identified.

Thirdly,
the [Fe/H] and [Eu/Fe] ratios of \jtwo\ 
are suggestive of a dwarf galaxy environment.
If \jtwo\ formed in an environment 
where an average \rpro\ event's yield was diluted 
into an average amount of Fe,
it would exhibit
[Fe/H] = $-2.20 + 1.04 = -1.16$ and
[Eu/Fe] = $+2.45 - 1.04 = +1.41$.
(We derived the value 1.04 in Section~\ref{sec:dilution}.)
While these ratios are rare among field stars,
they are found among several \rpro-enhanced stars
in the \fornax\ dwarf galaxy \citep{reichert21fnx}.

We explore this potential connection further by
expressing the relationship 
between the H, Fe, and Eu abundances
using standard abundance definitions
and the Solar abundance ratios:\
\begin{eqnarray}
\label{eqn:masseu1}
M_{\rm H} &=& M_{\rm Eu} \cdot 10^{9.30 - [{\rm Eu/H}]}, {\rm or} \\
\label{eqn:masseu2}
M_{\rm Eu} &=& M_{\rm H} \cdot 10^{[{\rm Eu/H}] - 9.30}.
\end{eqnarray}
Note that the mass of Eu ejected by an \rpro\ event ($M_{\rm Eu}$)
equals $\approx 10^{-3}$ times the
mass of \rpro\ elements ($M_{r}$) in a standard \rpro\ abundance pattern,
such as that found in the Solar System.
If the [Eu/H] ratio is known, and either of the
dilution mass of H ($M_{\rm H}$) or $M_{\rm Eu}$ can be estimated,
the third quantity can be calculated.
A comparison sample of six other stars 
with [Eu/Fe] $> +1.3$ listed in Table~\ref{tab:dilution}
exhibits a mean [Eu/H] of $-0.97 \pm 0.26$.
For a canonical baryonic minihalo mass of $M_{\rm H} = 10^{6}$~\msun\
(e.g., \citealt{tegmark97}),
Equation~\ref{eqn:masseu2} predicts an average yield of
$M_{\rm Eu} \approx 5\times10^{-5}$~\msun, or 
$M_{r} \approx 5\times10^{-2}$~\msun,
for the \rpro\ nucleosynthesis events that enriched 
the comparison sample.

Using this \rpro\ yield,
Equation~\ref{eqn:masseu1} predicts
\jtwo\ would have formed from
gas with $M_{\rm H} \approx 6\times10^{4}$~\msun,
assuming that the \rpro\ material is well mixed into the gas.
This mass would be even lower
if most of the \rpro\ material
was lost to the intergalactic medium,
rather than incorporated into stars.
\citet{beniamini18} estimated that a large fraction,
90\%, of \rpro\ material could be retained
in dwarf galaxies that had not already lost a 
large fraction of their gas.
The retainment fraction might also be a strong function
of where \rpro\ material is injected into a dwarf galaxy 
\citep{safarzadeh17ret2,tarumi20},
the timing of the \rpro\ enrichment \citep{bonetti19},
and the mass of the galaxy itself \citep{cavallo23}.
If only $\sim$~1\% of \rpro\ material was retained
in stars, 
as inferred from lighter metals in dwarf galaxies
(e.g., \citealt{kirby11out,mcquinn15,emerick18}),
then the \rpro\ material retained would have been diluted into 
only $\sim$~600~\msun\ of gas.
This gas cloud would have already been enriched by light elements
produced in normal ratios by typical core-collapse supernovae.
In this scenario,
very few stars with [Eu/Fe] ratios as high as that found in \jtwo\
could have formed.

\section{Summary and Conclusions}
\label{sec:conclusions}

\jtwo\ was observed
using the MIKE spectrograph at Magellan
as part of the high-resolution spectroscopic followup
of metal-poor star candidates 
conducted by the RPA.~
We immediately recognized the strong lines of 
lanthanides and other heavy elements in its spectrum
(Figures~\ref{fig:specplot}--\ref{fig:synth3}).
We derive stellar parameters (Table~\ref{tab:data})
that place this metal-poor star
([Fe/H] = $-2.20$)
on the red horizontal branch
(\teff\ = 5509~K, \logg\ = 2.28).
We calculate the kinematics of \jtwo,
and we find that it follows an eccentric orbit 
confined within the Solar circle
($r_{\rm peri}$ = 1.1~kpc, $r_{\rm apo}$ = 7.2~kpc).
We derive abundances of 47~metals detected in its optical spectrum,
and we derive upper limits on six other elements based on non-detections
(Table~\ref{tab:abund}).
The abundances of light elements ($Z \leq 30$) in 
\jtwo\ are typical for metal-poor halo stars.
The abundances of heavier elements, in contrast,
are highly enhanced
(e.g., [Eu/Fe] = $+2.45$), including many with super-Solar
[X/H] ratios
(e.g., [Eu/H] = $+0.25$),
and it is a member of the class of \rthree\ stars
that was proposed by \citet{cain20}.
The heavy-element abundance pattern is a close match to the
\rpro\ abundance pattern inferred for the Sun
(Figure~\ref{fig:rpropattern}) and
well-studied \rpro-enhanced stars
(Figure~\ref{fig:rstars}).
We detect radioactive isotopes of thorium and uranium,
and we calculate an age of $13.6 \pm 2.6$~Gyr 
for the \rpro\ material in \jtwo\ 
(Figure~\ref{fig:ages}, Table~\ref{tab:uranium}).

We investigate what may have led to the high level of
\rpro\ enhancement in \jtwo.
We propose a new method to assess the relative 
dilution or relative yields of \rpro\ material
incorporated into a star (Section~\ref{sec:dilution}).
This method makes use of
the correlation between the 
\logeps{Ru,Rh,Pd,Ag/Zr} ratios
and the [Eu/Fe] ratio
in metal-poor \rpro-enhanced stars
\citep{roederer23b}.
\jtwo\ exhibits \logeps{Ru,Rh,Pd,Ag/Zr} ratios
that are similar to other 
stars with [Eu/Fe] $> +1.3$,
suggesting that the yields of the \rpro\ events
that enriched these stars may have been similar.
We conclude that \jtwo\ 
exhibits a high level of \rpro\ enhancement
because it formed in an environment
where the \rpro\ material was less diluted than average.
If we assume a canonical baryonic minihalo mass of $10^{6}$~\msun\
and 1\% retention of \rpro\ material in star-forming gas,
then \jtwo\ would have formed where
$\approx 0.05$~\msun\ of \rpro\ material 
was diluted into $\approx 600$~\msun\ of gas.
This order-of-magnitude estimate requires 
several simplifying assumptions,
as discussed in Sections~\ref{sec:dilution} and \ref{sec:dwarfgalaxy},
but it may suggest
why stars like \jtwo\ are relatively uncommon.

\acknowledgments

In memory of 
Jim Lawler (1951--2023),
an uncommonly meticulous scientist and patient mentor.

We thank John Cowan for helpful feedback on an earlier version
of the manuscript, and we thank the anonymous referee for 
thoughful suggestions that have improved the manuscript.
We acknowledge support 
from the U.S.\ National Science Foundation (NSF):\
grants 
PHY~14-30152 (Physics Frontier Center/JINA-CEE),
OISE~1927130 
(International Research Network for Nuclear Astrophysics/IReNA),
AST~1716251 (A.F.), 
AST~1815403/1815767 and AST~2205847 (I.U.R.), 
and
AST-2206263 (R.E.).~
I.U.R.\ acknowledges support from the NASA
Astrophysics Data Analysis Program, grant 80NSSC21K0627.
K.H.\ is supported by JSPS KAKENHI Grant Numbers JP21K13965 and JP21H00053.
The work of V.M.P.\ is supported by NOIRLab, 
which is managed by AURA under a cooperative agreement with 
the NSF.~
This research has made use of NASA's
Astrophysics Data System Bibliographic Services;
the arXiv preprint server operated by Cornell University;
the SIMBAD and VizieR
databases hosted by the
Strasbourg Astronomical Data Center;
the ASD hosted by NIST;
and
the IRAF software packages
distributed by the National Optical Astronomy Observatories,
which are operated by AURA,
under cooperative agreement with the NSF.~
This work has also made use of data from the 
European Space Agency (ESA) mission Gaia 
(\url{https://www.cosmos.esa.int/gaia}), 
processed by the Gaia DPAC
(\url{https://www.cosmos.esa.int/web/gaia/dpac/consortium}). 
Funding for the DPAC
has been provided by national institutions, in particular the institutions
participating in the Gaia Multilateral Agreement.

\facility{Magellan~II (MIKE)}

\software{%
AGAMA \citep{vasiliev19},
IRAF \citep{tody93},
LINEMAKE \citep{placco21linemake},
MatPlotLib \citep{hunter07},
MOOG \citep{sneden73,sobeck11},
NumPy \citep{vanderwalt11},
R \citep{rsoftware},
SciPy \citep{jones01}}

\appendix
\restartappendixnumbering

\section{Ages Calculated from Radiogenic Lead Production}
\label{sec:appendix}

Following \citet{clayton64},
the abundance of each Pb isotope at time $t$, 
where $t$ is expressed in Gyr, is given by
\begin{eqnarray}
\label{eqn:pb1}
^{206}{\rm Pb}(t) 
&=& ^{206}{\rm Pb}_{r} 
  + ^{206}\!{\rm Pb}_{c} 
  + ^{206}\!{\rm Pb}_{s} \nonumber \\
^{207}{\rm Pb}(t) 
&=& ^{207}{\rm Pb}_{r} 
  + ^{207}\!{\rm Pb}_{c} 
  + ^{207}\!{\rm Pb}_{s} \\
^{208}{\rm Pb}(t) 
&=& ^{208}{\rm Pb}_{r} 
  + ^{208}\!{\rm Pb}_{c} 
  + ^{208}\!{\rm Pb}_{s}, \nonumber
\end{eqnarray}
where
\begin{eqnarray}
\label{eqn:pb2}
^{206}{\rm Pb}_{c} &=& ^{238}{\rm U}_{r} - ^{238}\!{\rm U}_{r}(t) \nonumber \\
^{207}{\rm Pb}_{c} &=& ^{235}{\rm U}_{r} - ^{235}\!{\rm U}_{r}(t) \\
^{208}{\rm Pb}_{c} &=& ^{232}{\rm Th}_{r} - ^{232}\!{\rm Th}_{r}(t) \nonumber
\end{eqnarray}
and
\begin{eqnarray}
\label{eqn:pb3}
^{238}{\rm U}_{r}(t)  &=& ^{238}{\rm U}_{r}  - ^{238}\!{\rm U}_{r} (1 - e^{-t\ln{2}/4.468}) \nonumber \\
^{235}{\rm U}_{r}(t)  &=& ^{235}{\rm U}_{r}  - ^{235}\!{\rm U}_{r} (1 - e^{-t\ln{2}/0.7038}) \\
^{232}{\rm Th}_{r}(t) &=& ^{232}{\rm Th}_{r} - ^{232}\!{\rm Th}_{r}(1 - e^{-t\ln{2}/14.0}). \nonumber 
\end{eqnarray}
In these equations,
the subscript \textit{r} denotes the direct isobaric \rpro\ contribution
and the contribution from short-lived translead or transuranic isotopes;
\textit{c} denotes the cosmoradiogenic decay of 
the long-lived $^{232}$Th, $^{235}$U, and $^{238}$U isotopes
in the time interval between nucleosynthesis and the present day;
and
\textit{s} denotes the \spro\ contribution.
The values of $^{A}{\rm X}_{r}$ are calculated theoretically
(e.g., \citealt{kratz07,roederer09b}).
The total Pb abundance is thus
\begin{eqnarray}
\label{eqn:pb4}
{\rm Pb} = ^{204}\!{\rm Pb} + 
^{206}\!{\rm Pb} + ^{207}\!{\rm Pb} + ^{208}\!{\rm Pb}.
\end{eqnarray}
The $^{204}$Pb isotope is blocked from \rpro\ production
by $^{204}$Hg.
We assume that the \spro\ contribution to the
Pb abundance in \jtwo\ is zero.
The Pb abundance at time $t$ is given by
\begin{eqnarray}
\label{eqn:pb5}
{\rm Pb}(t) = ^{206}\!{\rm Pb}_{r} +
              ^{207}\!{\rm Pb}_{r} + 
              ^{208}\!{\rm Pb}_{r} +
              ^{238}\!{\rm U}_{r} (1 - e^{-t\ln{2}/4.468}) +
              ^{235}\!{\rm U}_{r} (1 - e^{-t\ln{2}/0.7038}) + 
              ^{232}\!{\rm Th}_{r}(1 - e^{-t\ln{2}/14.0}).
\end{eqnarray}
Equation~\ref{eqn:pb5} is not easily solved analytically for $t$,
so we resort to numerical methods to estimate
$t$ from the observed ratios among the Pb, Th, and U abundances
in Section~\ref{sec:uranium}.

\bibliographystyle{aasjournal}
\bibliography{ms.bbl}

\end{document}